\documentclass[12pt,dvips]{article}
\usepackage[final]{graphics}
\usepackage{amssymb}

\newcommand{\cf}{cf.\ }

\newcommand{\eg}{e.g., }

\newcommand{\etal}{et al.}
\newcommand{\ie}{i.e., }

\newcommand{\sect}[1]{Section \ref{s:#1}}

\newcommand{\eqn}[1]{Eq.\ (\ref{e:#1})}
\newcommand{\eqntwo}[2]{Eqs.\ (\ref{e:#1}) and (\ref{e:#2})}
\newcommand{\eqns}[2]{Eqs.\ (\ref{e:#1})--(\ref{e:#2})}
\newcommand{\Eqn}[1]{Equation (\ref{e:#1})}

\newcommand{\fig}[1]{Fig.\ \ref{f:#1}}
\newcommand{\figtwo}[2]{Figs.\ \ref{f:#1} and \ref{f:#2}}

\newcommand{\Fig}[1]{Figure \ref{f:#1}}

\newcommand{\code}[1]{\texttt{#1}}

\newcommand{\bvec}[1]{\mathbf{#1}}           
\newcommand{\bhat}[1]{\bvec{\hat{#1}}}       
\newcommand{\bmat}[1]{\bvec{#1}}             
\newcommand{\bsym}[1]{\mbox{\boldmath $#1$}} 
\newcommand{\vdot}{\bsym{\cdot}}             
\newcommand{\cross}{\bsym{\times}}           
\newcommand{\trans}[1]{#1^\mathrm{T}}        

\def\paper#1 #2 #3 #4 #5 #6 {#1, #2. #3. #4\ #5, #6.}
\def\inpress#1 #2 #3 #4 {#1, #2. #3. #4, in press.}
\def\preprint#1 #2 #3 #4 {#1, #2. #3. ArXiv e-prints #4.}
\def\submitted#1 #2 #3 #4 {#1, #2. #3. #4, submitted.}
\def\inprep#1 #2 #3 #4 {#1, #2. #3. #4, in preparation.}
\def\thesis#1 #2 #3 #4 #5 {#1, #2. #3. Thesis, #4. #5 pp.}
\def\book#1 #2 #3 #4 {#1, #2. #3. #4.}
\def\chap#1 #2 #3 #4 #5 #6 #7 {#1, #2. #3. In: #4 (Eds.), #5. #6, pp.\ #7.}
\def\aIII#1 #2 #3 {#1, 2002. #2. In: Bottke Jr., W.F., Cellino, A.,
  Paolicchi, P., Binzel, R.P. (Eds.), Asteroids III. Univ.\ of Arizona
  Press, Tucson, pp.\ #3.}

\newcommand{\hide}[1]{} 

\newcounter{fignum}

\newcommand{\putfig}[2]{ 
  \clearpage
  \refstepcounter{fignum}
  \label{f:#1}
  \begin{center}
    Figure \thefignum
    \vfill
    \resizebox{#2}{!}{\includegraphics{#1.ps}}
    \vfill
  \end{center}
}

\newcommand{\figcap}[1]{\item{\bfseries Figure \ref{f:#1}:}}

\begin{document}
%
%
\setlength{\footskip}{0pt} 
\begin{center}
  \vspace*{\fill}
  \textbf{\large Numerical simulations of granular
    dynamics. I. Hard-sphere discrete element method and tests}\\
  \vfill
  \textbf{Derek C. Richardson}$^\dag$\\
  \textbf{Kevin J. Walsh}$^\ddag$\\
  \textbf{Naomi Murdoch}$^{\ddag\star}$\\
  \textbf{Patrick Michel}$^\ddag$\\
  \bigskip
  $^\dag$Department of Astronomy\\
  University of Maryland\\
  College Park MD 20740-2421\\
  \bigskip
  $^\ddag$University of Nice Sophia Antipolis, CNRS\\
  Observatoire de la C\^ote d'Azur, B.P. 4229\\
  06304 Nice Cedex 4, France\\
  \bigskip
  $^\star$The Open University\\
  PSSRI, Walton Hall\\
  Milton Keynes, MK7 6AA, UK\\
  \bigskip
  Printed \today\\
  \bigskip
  Submitted to \textit{Icarus}\\
  \vfill
  54 manuscript pages\\
  8 figures including 4 in color (online version only)\\
  \vfill
\end{center}

\newpage
\begin{flushleft}
  Proposed running page head: Granular dynamics method and tests\\
  \bigskip
  Please address all editorial correspondence and proofs to:\\
  \bigskip
  Derek C. Richardson\\
  Department of Astronomy\\
  Computer and Space Sciences Building\\
  Stadium Drive\\
  University of Maryland\\
  College Park MD 20742-2421\\
  Tel: 301-405-8786\\
  Fax: 301-314-9067\\
  E-mail: \code{dcr@astro.umd.edu}
\end{flushleft}

\newpage
\section*{Abstract}

We present a new particle-based (discrete element) numerical method
for the simulation of granular dynamics, with application to motions
of particles on small solar system body and planetary surfaces.  The
method employs the parallel $N$-body tree code \code{pkdgrav} to
search for collisions and compute particle trajectories.  Collisions
are treated as instantaneous point-contact events between rigid
spheres.  Particle confinement is achieved by combining arbitrary
combinations of four provided wall primitives, namely infinite plane,
finite disk, infinite cylinder, and finite cylinder, and degenerate
cases of these.  Various wall movements, including translation,
oscillation, and rotation, are supported.  We provide full derivations
of collision prediction and resolution equations for all geometries
and motions.  Several tests of the method are described, including a
model granular ``atmosphere'' that achieves correct energy
equipartition, and a series of tumbler simulations that show the
expected transition from tumbling to centrifuging as a function of
rotation rate.

\begin{description}
\item{\textbf{Keywords}: Asteroids, surfaces; Collisional physics;
  Geological processes; Regoliths} 
\end{description}

%
%
\newpage
\section{INTRODUCTION} \label{s:intro}

The understanding of granular dynamics is taking an increasingly
prominent place in the field of planetary science.  Indeed, we now
know that granular material, in the form of regolith, is covering the
uppermost layer of most solid bodies in our solar system, from planets
and their satellites to asteroids and comets.  This presence of a
regolith layer plays a particularly important role in the surface
geology of asteroids.  The same can be stated, although to a lesser
extent, for bodies like Mars and the Moon, whose surface gravities are
also smaller than that of Earth.  Thus, flows of granular materials
driven by different gravitational conditions are particularly
important in the understanding of the geology of small bodies and
planets.  The exploration of the Moon and Mars in the next two decades
will require deployment of landing vehicles on surfaces of loose
granular material.  Space missions to small bodies also involve
measurements by landers (\eg the Rosetta space mission of the European
Space Agency) and sampling devices capable of coping with a wide range
of surface properties.  Therefore, understanding how granular
materials, as a function of their properties (angle of friction, size
distribution of their components, etc.), react to different kinds of
stresses is of great interest for the design of landers and sampling
devices of space missions.  The same holds true in the framework of
mitigation strategies against a potential impactor, which involve an
interaction with the body's surface.

The presence or relative absence of gravitational acceleration on
granular flow is of importance for understanding the geology of small
bodies and planets, and to clarify the environments that may be
encountered during planetary exploration.  Bodies with low surface
gravity can be very sensitive to processes that appear irrelevant in
the case of larger planetary bodies.  For instance, seismic vibration
induced by small impacts can occur throughout a small body and can be
at the origin of motion of its granular surface.  Such a mechanism has
been proposed to explain the lack of very small craters on both
asteroid 433 Eros (Richardson \etal\ 2004) and asteroid 25143 Itokawa
(Michel \etal\ 2009).  Shaking can also drive size sorting/segregation
in granular media.  This has been observed on Earth (Rosato
\etal\ 1987), as well as on Eros (Robinson \etal\ 2002) and Itokawa
(Miyamoto \etal\ 2007).  In particular, on asteroid Itokawa it has
been observed that the neck is formed mainly by centimeter or
smaller-sized regolith but that both the head and body are covered by
meter-scale boulders.  This finding has led some to suggest that
subsequent impacts on an asteroid could provide the energy and driving
mechanism for segregation to occur (Asphaug \etal\ 2001, Miwa
\etal\ 2008), leading to a phenomenon commonly called the ``Brazil-nut
effect.''  However, it was then found for Itokawa that an erosion
mechanism was not sufficient to explain the selective location of
coarse material in the potential lows and highs, and that other
factors, such as cohesion, rotation, or avalanches may be involved
(Sanchez \etal\ 2010).  In fact, rotational spin-up of a small body as
a result of the thermal YORP effect can lead to regolith motion on the
surface (Scheeres \etal\ 2007), and even to the escape of material
that can eventually result in the formation of a satellite (Walsh
\etal\ 2008).

All these aspects demonstrate the great importance of understanding
the dynamics of granular matter for planetary science applications.
The study of granular dynamics requires dealing with complex
mechanical processes and currently constitutes an entire field of
research by itself.  As examples of the complexity, it has been found
that granular matter may strengthen under a variety of conditions, for
instance as a result of a slow shift in the particle arrangement under
shear stress or because of humidity (Losert \etal\ 2000b).  Moreover,
while granular materials are a discrete medium, their flow in response
to stress can, under certain conditions, be described by continuum
models (Savage 1998; Goddard 1990; Losert \etal\ 2000).  However, many
processes cannot be captured using classical continuum approaches.
For example, highly stressed granular materials exhibit shear
localization during failure (Mueth \etal\ 2000) but this localization
cannot be described accurately by existing continuum models (Kamrin
\etal\ 2007).  In addition, situations have been identified in which
material under shear stress weakens when the shear direction is
changed/reversed (Toiya \etal\ 2004; Falk \etal\ 2008).  Finally, at
low gravity, other physical forces relevant to regolith in the
asteroid environment, such as van der Waals forces, may exceed the
particle weights and require consideration (Scheeres \etal\ 2010).

Improving our understanding of the dynamics of granular materials
under a wide variety of conditions requires that both experimental and
numerical work be performed and compared.  Once numerical approaches
have been validated by successful comparison with experiments, then
they can cover a parameter space that is too wide for or unreachable
by laboratory experiments.

There are several different approaches that have been used to perform
modeling of granular materials (Mehta 2007).  One commonly used
technique is the discrete element method (DEM).  DEM is a numerical
method for computing the motion of a large number of particles of
micron-scale size and above.  Though DEM is very closely related to
molecular dynamics (in which atoms and molecules are allowed to
interact for a period of time by approximations of known physics), the
method is generally distinguished by its inclusion of rotational
degrees of freedom, inelastic collisions, and often-complicated
geometries (including polyhedra; \eg Cleary and Sawley 2002; Fraige
\etal\ 2008; Latham \etal\ 2008; Szarf \etal\ 2010).

However, the DEM method remains relatively computationally intensive,
which limits either the length of a simulation or the number of
particles.  Several DEM codes, and related molecular dynamics codes,
take advantage of parallel processing capabilities to scale up the
number of particles or length of the simulation (\eg Cleary and Sawley
2002; Kacianauskas \etal\ 2008).

Hard-sphere particle dynamics have been used successfully in many
granular physics applications.  Hong and McLennan (1992) used
hard-sphere molecular dynamics to study particles flowing through a
hole in a two-dimensional box under the influence of gravity.  Huilin
\etal\ (2007) used an Eulerian-Lagrangian approach coupled with a
discrete hard-sphere model to obtain details of particle collision
information in a fluidized bed of granular material.  Also Kosinski
and Hoffman (2009) compared the standard hard-sphere method including
walls to a hard-sphere model with walls that also accounts for
particle adhesion.  The van der Waals type interaction is presented as
a demonstration case.

An alternative to treating all particles separately is to average the
physics across many particles and thereby treat the material as a
continuum.  In the case of solid-like granular behavior, the continuum
approach usually treats the material as elastic or elasto-plastic and
models it with the finite element method or a mesh-free method (\eg
Elaskar \etal\ 2000; also see Holsapple 2001 and 2004, Holsapple and
Michel 2006 and 2008, and Sharma \etal\ 2006 and 2009 for use of
analytical and continuum approaches in modeling asteroid shapes).  In
the case of liquid-like or gas-like granular flow, the continuum
approach may treat the material as a fluid and use computational fluid
dynamics.  However, as explained previously, the homogenization of the
granular-scale physics is not necessarily appropriate, and the
discreteness of the particles and the forces between particles (and
walls) need to be taken into account (Wada \etal\ 2006).  Therefore
limits of such homogenization must be considered carefully before
attempting to use a continuum approach.

Recently, numerical codes have been developed to address specifically
the dynamics of granular materials in the framework of planetary
science.  For instance, Sanchez \etal\ (2010) simulated the particles
forming an asteroid by means of soft-sphere molecular dynamics.  In
this approach, particles forming the aggregate have short- and
long-range interactions, for contact and gravitational forces
respectively, which are taken into account using two types of
potentials (see Sanchez \etal\ 2010 for details).

In this paper, we present our method to simulate the dynamics of
granular materials, along with a few basic tests that show its ability
to address different problems involving these dynamics.  A more
complete validation by comparison with a well-documented laboratory
experiment of seismic shaking is the subject of a follow-up paper
(Murdoch \etal\ 2011, in preparation).  We use the $N$-body code
\code{pkdgrav} (Stadel 2001), adapted for hard-body collisions
(Richardson \etal\ 2000; Richardson \etal\ 2009).  The granular
material is therefore represented by hard spheres that interact via
impulsive, point-contact forces.  Advantages of \code{pkdgrav} over
other many discrete element approaches include full support for
parallel computation, the use of hierarchical tree methods to rapidly
compute long-range interparticle forces (namely gravity, when
included) and to locate nearest neighbors (for short-range Hooke's-law
type forces) and potential colliders, and options for particle bonding
to make irregular shapes that are subjet to Euler's laws of rigid-body
rotation with non-central impacts (\cf Richardson \etal\ 2009).  In
addition, collisions are determined prior to advancing particle
positions, insuring that no collisions are missed and that collision
circumstances are computed exactly (in general, to within the accuracy
of the integration), which is a particular advantage when particles
are moving rapidly.

Here we focus on the implementation of a wide range of boundary
conditions (``walls'') that can be used to represent the different
geometries involved in experimental setups, but more generally provide
the needed particle confinement and possible external forcings, such
as induced vibrations, for small-scale investigations of regolith
dynamics in varying gravitational environments (different surface
slopes, etc.).  Our approach is designed to be general and flexible:
any number of walls can be combined in arbitrary ways to match the
desired configuration without changing any code, whereas many existing
methods are tailored for a specific geometry.  The long-term goal is
to understand how scaling laws, different flow regimes, segregation,
and so on change with gravity, and to apply this understanding to
asteroid surfaces, without the need to simulate the surfaces in their
entirety.  We present the full detail of our numerical method in
\sect{method}, describe basic tests of the method in \sect{tests}, and
offer discussion and conclusions in \sect{disc} and \sect{concl},
respectively.

\section{NUMERICAL METHOD} \label{s:method}

This section details the numerical method employed in our granular
dynamics simulations.  We use \code{pkdgrav}, a parallel $N$-body tree
code (Stadel 2001) that has been adapted for hard-body collisions
(Richardson \etal\ 2000; Richardson \etal\ 2009).  In what follows we
present the general strategy of searching for and resolving collisions
among particles, and between particles and the ``walls'' of the
simulated apparatus.  This is followed by detailed derivations of the
collision equations for the four wall ``primitives'' that we have
developed so far, namely the infinite plane, finite disk, infinite
cylinder, and finite cylinder, plus degenerate cases of these.

\subsection{General Strategy} \label{s:strategy}

In our approach, we predict when collisions will occur based on
trajectory extrapolation at the beginning of each integration step.
We use a second-order leapfrog integrator in \code{pkdgrav}: each step
consists of ``kicking'' particle velocities by a half step (keeping
particle positions fixed), ``drifting'' particle positions at constant
velocity for a full step, recomputing accelerations due to gravity,
then performing a final half-step velocity kick (see Richardson
\etal\ 2009 for details).  Collision searches are performed during the
drift step by examining the trajectories of enough neighbors of each
particle to ensure no collisions are missed.  Since there is no
interparticle gravity for the granular dynamics experiments discussed
here (only a configurable uniform gravity field), the equations of
motion are particularly simple, but we nonetheless use the full
machinery of the $N$-body code for these simulations---the tree is
still used to speed up neighbor searches, and parallelism reduces
computation time for large simulations.  Also, more complex gravity
fields for which there is no analytical solution for particle motion
will be used in the future.

The condition for two particles with initial vector positions
$\bvec{r}_1$ and $\bvec{r}_2$ and velocities $\bvec{v}_1$ and
$\bvec{v}_2$ to collide after a time interval $t$ (measured from the
start of the drift step in this case) is
\begin{equation}
  |\bvec{r}_2 - \bvec{r}_1 + (\bvec{v}_2 - \bvec{v}_1) t| = s_1 + s_2
  , \label{e:sphcolcomp}
\end{equation}
where $s_1$ and $s_2$ are the radii of the particles (treated as
perfect hard spheres), or
\begin{equation}
  |\bsym{\rho} + \bsym{\nu} t| = s_1 + s_2 , \label{e:sphcol}
\end{equation}
where we have introduced the relative position $\bsym{\rho} \equiv
\bvec{r}_2 - \bvec{r}_1$ and velocity $\bsym{\nu} \equiv \bvec{v}_2 -
\bvec{v}_1$ (\fig{spheres.pptx}).  To collide, the particles must
be approaching one another, so $\bsym{\rho} \vdot \bsym{\nu} < 0$.  We
assume the particles are not initially touching or overlapping, \ie we
require $\rho > s_1 + s_2$, where $\rho \equiv |\bsym{\rho}|$.
\Eqn{sphcol} can be solved for $t$ by squaring both sides and applying
the quadratic formula:\footnote{In practice, a version of the
  quadratic formula optimized to reduce round-off error is used in
  \code{pkdgrav}---\cf Press \etal\ 2007.}
\begin{equation}
  t = \frac{- (\bsym{\rho} \vdot \bsym{\nu}) \pm \sqrt{(\bsym{\rho}
      \vdot \bsym{\nu})^2 - \left[\rho^2 - (s_1 + s_2)^2\right]
      \nu^2}}{\nu^2} , \label{e:twoparticles}
\end{equation}
where $\nu \equiv |\bsym{\nu}|$ and the sign ambiguity is resolved by
taking $t$ to be the smallest positive value of any real roots.  If
$t$ is between zero and the drift step time interval (\ie the
simulation time step), the collision takes place during this
integration step.  Collisions are treated as instantaneous events with
no flexing and a configurable amount of energy loss due to restitution
(parameterized by the normal coefficient of restitution $0 \le
\varepsilon_n \le 1$, where 0 means perfect sticking and 1 means
perfect bouncing) and surface coupling (parameterized by the
transverse coefficient of restitution $-1 \le \varepsilon_t \le 1$,
where $-1$ means reversal of transverse motion on contact, 0 means
complete damping of transverse motion, and 1 means no surface
coupling).  For completeness, the collision resolution equations for
perfect spheres are (Richardson \etal\ 2009, Appendix B; also see
Richardson 1994, Fig.\ 1):
\begin{eqnarray*}
  \bvec{v}_1^\prime & = & \bvec{v}_1 + \frac{m_2}{M} \left[ (1 +
    \varepsilon_n) \bvec{u}_n + \frac{2}{7} (1 - \varepsilon_t)
    \bvec{u}_t \right] , \\
  \bvec{v}_2^\prime & = & \bvec{v}_2 - \frac{m_1}{M} \left[ (1 +
    \varepsilon_n) \bvec{u}_n + \frac{2}{7} (1 - \varepsilon_t)
    \bvec{u}_t \right] , \\
  \bsym{\omega}_1^\prime & = & \bsym{\omega}_1 + \frac{2}{7}
  \frac{\mu (1 - \varepsilon_t)}{I_1} (\bvec{s}_1 \cross \bvec{u}) ,
  \\
  \bsym{\omega}_2^\prime & = & \bsym{\omega}_2 - \frac{2}{7}
  \frac{\mu (1 - \varepsilon_t)}{I_2} (\bvec{s}_2 \cross \bvec{u}) ,
\end{eqnarray*}
where the primes denote post-collision quantities, $M = m_1 + m_2$ is
the sum of the particle masses, $\bvec{u} = \bsym{\nu} +
(\bsym{\sigma}_2 - \bsym{\sigma}_1)$ is the total relative velocity at
the contact point (with $\bsym{\sigma}_i = \bsym{\omega}_i \cross
\bvec{s}_i$, $i = 1,2$, where $\bsym{\omega}_i$ is the spin vector of
particle $i$, $\bvec{s}_i = (-1)^{(i-1)} s_i \bhat{n}$, and $\bhat{n}
= \bsym{\rho}/\rho$), $\bvec{u}_n = (\bvec{u} \vdot \bhat{n})
\bhat{n}$, $\bvec{u}_t = \bvec{u} - \bvec{u}_n$, $\mu$ is the reduced
mass $m_1 m_2/M$, and $I_i = \frac{2}{5} m_i s_i^2$ is the moment of
inertia of particle $i$; the factors of 2/7 result from using spheres.
These equations come from conservation of linear and angular momentum
combined with the following statement of energy loss:
\begin{equation}
  \bvec{u}^\prime = -\varepsilon_n \bvec{u}_n + \varepsilon_t
  \bvec{u}_t .
\end{equation}

\begin{center}
  [FIGURE \ref{f:spheres.pptx} GOES HERE]
\end{center}

The approach for handling wall collisions is similar: first the time
to collision is determined, then the collision is resolved.  The
collision condition is that the distance of the particle center
from the point of contact on the wall must equal the particle radius,
\ie
\begin{equation}
  |\bvec{r}_{\mathrm{impact}} - \bvec{c}| = s ,
\end{equation}
where $\bvec{c}$ is the vector position of the point of contact and we
have dropped the subscript $i$.  (For the derivations that follow, we
have adopted the convention that the wall corresponds to $i = 1$ and
the particle to $i = 2$; this minimizes the number of minus signs in
the equations.)  Because $\bvec{c}$ depends on the particular wall
geometry, it is best to consider each geometry in turn and exploit any
symmetries to determine more conveniently when (if at all) this
contact condition is satisfied for a given particle and wall pair.
Note that often a wall consists of a combination of more than one
geometry (for example, a finite disk is a round portion of a plane
plus a surrounding ring, \ie a cylinder of zero length)---the
collision prediction routines for walls in \code{pkdgrav} return a
list of possible collision times, one for each ``face'' (for the
finite disk example, there are 3 faces: one flat side, the opposite
flat side, and the perimeter ring), with the smallest positive time
corresponding to the face that will be struck first.

In \code{pkdgrav}, after all neighbors of a given particle have been
checked for possible collisions, every wall is checked also.  Whatever
body (whether particle or wall) that gives the shortest time to
collision is taken to be the next collision event.  That collision is
resolved, collision times are updated depending on whether the
collision that just happened changes future collision circumstances,
and the next collision is carried out until all collisions during the
current drift interval have been handled.  Walls are treated as having
infinite mass, so the restitution equations reduce to:
\begin{eqnarray}
  \bvec{v}^\prime & = & \bvec{v} - (1 + \varepsilon_n) \bvec{u}_n -
  \frac{2}{7} (1 - \varepsilon_t) \bvec{u}_t , \label{e:wallbounce1}
  \\
  \bsym{\omega}^\prime & = & \bsym{\omega} - \frac{5}{7} s^{-2} (1 -
  \varepsilon_t) \bvec{s} \cross \bvec{u} , \label{e:wallbounce2}
\end{eqnarray}
where
\begin{equation}
  \bvec{u} = \bvec{v} + \bsym{\omega} \cross \bvec{s}
  \label{e:staticwall}
\end{equation}
(for a static wall), $\varepsilon_n$ and $\varepsilon_t$ are specific
to the wall (see below), and $\bvec{s} \equiv - s \bhat{n}$, with
$\bhat{n}$ being the unit vector in direction $(\bvec{r} + \bvec{v}t)
- \bvec{c}$, \ie from the point of contact (which depends on the wall
geometry) to the particle center.  To account for any wall motion,
$\bvec{u}$ in \eqn{staticwall} is adjusted as needed (see below).

The parameters of each wall type are specified by the user at run time
via a simple text file.  Information common to each wall that must be
specified includes the origin (a vector reference point), orientation
(a unit vector), $\varepsilon_n$ and $\varepsilon_t$ for the wall
(which override the particle values during a particle-wall collision),
and color and transparency (for drawing purposes).  In addition, each
geometry has a unique set of parameters particular to that geometry,
\eg radius in the case of the finite disk and cylinders, and length
for the finite cylinder.  Also, simple motions are supported for each
geometry; \eg a plane or disk can have translational and/or sinusoidal
oscillatory motion, while a cylinder can rotate around its symmetry
(orientation) axis.  Oscillations are treated by updating the wall
position in stepwise fashion, holding the velocity constant between
timesteps.  In \code{pkdgrav}, a single function is used to get the
position and velocity of a particle relative to any (possibly moving)
wall at a given time.  Finally, a wall can be ``sticky,'' meaning any
particles that come into contact with it become rigidly held, or
``absorbing,'' meaning any particles that come into contact with it
are removed from the simulation.  These latter options are encoded as
special cases of the wall normal coefficient of restitution (0
indicating sticky, $< 0$ indicating absorbing).  The sticky wall is
particularly useful for creating a rough surface (\cf \sect{tumbler});
the absorbing wall is helpful when particles are no longer needed,
such as when they have moved beyond the flow regime of interest.

For a particle hitting another particle that is rigidly stuck to a
wall, \eqntwo{wallbounce1}{wallbounce2} also apply when resolving the
collision (the stuck particle is treated as having infinite mass),
with $\bhat{n}$ pointing from the center of the stuck particle to the
center of the free particle, and $\varepsilon_n$ and $\varepsilon_t$
being the usual particle values (\ie not the values specific to the
wall).  Detection of collisions with stuck particles is handled by
setting the stuck particles' velocities equal to the instantaneous
velocities of their host walls at the start of the step.  In the case
of particles stuck on rotating cylinders, the component $\bvec{\Omega}
\cross \left[ (R \pm s) \bhat{n} \right]$ is added to the particle
velocities, where $\bvec{\Omega}$ is the cylinder's rotation vector
and $R$ is its radius, and in this case $\bhat{n}$ is the
perpendicular from the cylinder rotation axis to the particle center
at the start of the step (the sign of $s$ depends on whether the
particle is on the outer [positive] or inner [negative] surface).  The
positions of particles stuck on moving/rotating walls are updated at
the end of the drift step.  Collisions with particles stuck on
rotating cylinders can optionally be predicted to higher order using a
quartic expression that accounts for the curvilinear motion (\cf
Richardson \etal\ 2009, Eq.\ (A.4)).

For resolving collisions with moving walls, and particles stuck to
moving walls, each wall's instantaneous velocity at the start of the
step is substracted from $\bvec{u}$ in \eqn{staticwall} before
applying \eqntwo{wallbounce1}{wallbounce2}.  For cylinder rotation,
with the rotation axis parallel to the symmetry axis of the cylinder,
$\bvec{\Omega} \cross (R \bhat{n})$ is subtracted, where $\bhat{n}$ is
as defined for those equations.  For particles stuck to rotating
walls, the stuck particle itself is given an extra speed component
(see above), which is taken into account when resolving the collision
outcome.

\subsection{Specific Geometries}

The following sections detail the implementation of the supported wall
geometries in \code{pkdgrav}.

\subsubsection{Infinite plane} \label{s:infiniteplane}

The parameters for the infinite plane are the origin $\bvec{O}$ and
normal $\bhat{N}$, plus optional velocity $\bvec{V}$, oscillation
amplitude $A$, and oscillation angular frequency $\Omega$ (so the
relative vector displacement after time $T$ due to oscillation,
measured from the start of the simulation and evaluated at the start
of the step, is $A \sin(\Omega T) \bhat{N}$).  The origin can be any
point in the plane (the choice is arbitrary).

To simplify the equations in this and subsequent derivations, we
define the relative position vector $\bsym{\rho} \equiv \bvec{r} -
\bvec{O}$ and separate it into perpendicular and parallel components,
$\bsym{\rho}_N$ and $\bsym{\rho}_T$, respectively, where
$\bsym{\rho}_N \equiv \rho_N \bhat{N}$, $\rho_N \equiv \bsym{\rho}
\vdot \bhat{N}$, and $\bsym{\rho}_T \equiv \bsym{\rho} -
\bsym{\rho}_N$ (so $\bhat{T} \equiv \bsym{\rho}_T/|\bsym{\rho}_T|$,
which is only defined if $|\bsym{\rho}_T| > 0$ and means $\bhat{T}$
generally has a time dependence; for completeness, we also define
$\rho_T \equiv \bsym{\rho} \vdot \bhat{T}$).  It is important to note
that $\rho_N$ and $\rho_T$ are \emph{signed} quantities (vector
components), unlike $\rho$ defined previously, which is an unsigned
magnitude.  We similarly define the relative velocity $\bsym{\nu}
\equiv \bvec{v} - \dot{\bvec{O}}$, with corresponding perpendicular
and parallel components.

Returning to the specific case of the infinite plane, the collision
condition is $|\rho_{N,\mathrm{impact}}| = s$, \ie that the
perpendicular distance from the particle center to the surface (the
height above or below the plane) at the time of impact is equal to the
particle radius (\fig{plane.pptx}).  For this geometry, and
measured from the start of the drift step, the time to impact is
\begin{equation}
  t = \left\{
  \begin{array}{ll}
    \frac{s - \rho_N}{\nu_N} & \mbox{if $\rho_N > 0$} , \\
    - \frac{s + \rho_N}{\nu_N} & \mbox{if $\rho_N < 0$} ,
  \end{array}
  \right. \label{e:planeimpact}
\end{equation}
where the first condition corresponds to impact with the ``upper''
face (the surface out of which the normal vector $\bhat{N}$ points)
and the second corresponds to the opposite ``lower'' face.  (We take
this two-case approach to avoid having to solve a quadratic; \cf
\sect{infinitecylinder}.)  Note $\rho_N \nu_N < 0$ is a requirement
for collision (otherwise the particle is moving away from or parallel
to the plane).  For completeness, the wall position at the start of the
step is $\bvec{O} = \bvec{O}(0) + \bvec{V}T + A \sin(\Omega T)
\bhat{N}$, where $\bvec{O}(0)$ is the origin at the start of the
simulation, and the wall velocity is $\dot{\bvec{O}} = \bvec{V} +
\Omega A \cos(\Omega T) \bhat{N}$.

\begin{center}
  [FIGURE \ref{f:plane.pptx} GOES HERE]
\end{center}

To resolve the collision, \eqntwo{wallbounce1}{wallbounce2} apply,
with $\bhat{n}$ being either $\bhat{N}$ or $-\bhat{N}$, depending on
which face was struck; specifically, $\bhat{n} =
\mathrm{sign}\left[\rho_{N,\mathrm{impact}}\right] \bhat{N}$.

\subsubsection{Finite disk} \label{s:finitedisk}

This geometry uses the same parameters as the infinite plane
(\fig{plane.pptx}) but now includes the finite radius $R$ of the
object.  The origin $\bvec{O}$ is the geometric center of the disk.
Collision prediction proceeds in 3 stages: first, \eqn{planeimpact} is
used to determine the impact time (if any) with the infinite plane
that contains the disk (\ie taking $R = \infty$); second, the position
the particle would have relative to the (possibly moving) disk at that
impact time is checked to determine whether
$|\rho_{T,\mathrm{impact}}| < R$, \ie that the separation of the
particle center from the disk origin, projected onto the plane, would
be less than the disk radius, indicating actual contact with the flat
surface of the finite disk; third, the impact time (if any) with the
disk perimeter, \ie a ring, is checked (see \sect{ring})---the
smallest positive time for impact with both faces and the perimeter is
taken to be the next impact time with this object.

Collision resolution is the same as for the infinite plane, unless the
particle has made contact with the perimeter, in which case the
collision is treated as a ring bounce (\sect{ring}).

\subsubsection{Infinite cylinder} \label{s:infinitecylinder}

The parameters for the infinite cylinder are the origin $\bvec{O}$,
symmetry-axis orientation $\bhat{N}$, radius $R$, and optional angular
frequency $\Omega$ (with $\bhat{N}$ as the spin axis).  The origin is
any point along the symmetry axis.  In the current implementation,
cylinders are fixed in space, and since the rotation doesn't change
the orientation (it only affects the tangential motion and spin of
particles that come into contact with the cylinder, assuming
$\varepsilon_t \neq 1$), $\bvec{O}$ is constant (so $\dot{\bvec{O}} =
0$).  The collision condition is $|\rho_{T,\mathrm{impact}}| = R \pm
s$, \ie the perpendicular distance from the particle center to the
cylinder wall is equal to the particle radius.  Note that particles
can either be inside (assuming $R > s$) or outside the cylinder; each
face is considered in turn.  \Fig{cylinder.pptx} illustrates the
geometry.

\begin{center}
  [FIGURE \ref{f:cylinder.pptx} GOES HERE]
\end{center}

To find the collision time (if any), note
$\bsym{\rho}_{\mathrm{impact}} = \bsym{\rho} + \bsym{\nu} t$, where
recall $\bsym{\rho} \equiv \bvec{r} - \bvec{O}$, $\bsym{\nu} \equiv
\bvec{v} - \dot{\bvec{O}} = \bvec{v}$ in this case, and $t$ is the
impact time measured from the start of the drift step.  Substituting
into the collision condition equation,
\begin{equation}
  \left| \bsym{\rho}_T + \bsym{\nu}_T t \right| = R \pm s
  . \label{e:cylinder}
\end{equation}
Taking the square,
\begin{equation}
  \rho_T^2 + 2 (\bsym{\rho}_T \vdot \bsym{\nu}_T) t + \nu_T^2 t^2 = (R
  \pm s)^2 .
\end{equation}
This is a quadratic equation for $t$ and can be solved in the usual
way.  For completeness,
\begin{equation}
  t = \frac{- (\bsym{\rho}_T \vdot \bsym{\nu}_T) \pm \sqrt{
      (\bsym{\rho}_T \vdot \bsym{\nu}_T)^2 - \nu_T^2
      \left[ \rho_T^2 - (R \pm s)^2 \right]}}{\nu_T^2}
  . \label{e:cylinderimpact}
\end{equation}
Note there are \emph{four} solutions, two each for impacts with the
inner face (corresponding to $R - s$ on the right-hand side of
\eqn{cylinder}) and outer face ($R + s$).  Again, for each face, the
smallest positive (real) value for $t$ is chosen.  If $\nu_T = 0$,
there will be no impact (the particle is either stationary with
respect to the cylinder, or moving parallel to the cylinder's symmetry
axis).  If the particle is outside the cylinder (so $|\rho_T| > R +
s$), a collision can only occur if $\bsym{\rho}_T \vdot \bsym{\nu}_T <
0$; for an infinite cylinder, the collision can only be with the outer
face.  If the particle is inside the cylinder (so $|\rho_T| < R - s$),
all solutions are real (\ie there must be a collision, assuming $\nu_T
\ne 0$); furthermore, only collisions with the inner face are possible
(even in the case of a finite cylinder).  Together these facts can be
used to reduce the computation time involved with this geometry.  Note
however, unlike other collision circumstances handled by
\texttt{pkdgrav}, in the case of an interior bounce involving a
cylinder (or a ring), it is possible for the particle to hit the same
object two or more times in one integration step without an
intervening collision with another object.

To resolve the collision, \eqntwo{wallbounce1}{wallbounce2} apply,
with $\bhat{n}$ being either $\bhat{T}_{\mathrm{impact}}$, if the
outer face was struck (requiring $|\rho_{T,\mathrm{impact}}| = R +
s$), or $-\bhat{T}_{\mathrm{impact}}$, if the inner face was struck
(requiring $|\rho_{T,\mathrm{impact}}| = R - s$).  Recall a cylinder
can be spinning, in which case $\Omega \bhat{N} \cross
\bsym{\rho}_{T,\mathrm{impact}}$ is subtracted from the total relative
velocity $\bvec{u}$ in \eqn{staticwall} before applying the collision
equations.

\subsubsection{Finite cylinder} \label{s:finitecylinder}

This geometry uses the same parameters as the infinite cylinder, but
now includes the finite length $L$ of the object.  The origin
$\bvec{O}$ is the geometric center of the cylinder.  In a manner
similar to the finite disk case (\sect{finitedisk}), collision
prediction proceeds in 3 stages: first, \eqn{cylinderimpact} is used
to determine the impact time (if any) with the infinite counterpart to
the cylinder (\ie taking $L = \infty$); second, the position the
particle would have relative to the cylinder at that impact time is
checked to determine whether $|\rho_{N,\mathrm{impact}}| < L/2$, \ie
that the separation of the particle center from the cylinder origin,
projected onto the cylinder's orientation axis, would be less than
half the cylinder length, indicating actual contact with the surface
of the finite object (not the ends); third, the impact times (if any)
with the cylinder ends are checked (see \sect{ring})---the smallest
positive time for impact with both faces and the cylinder ends is
taken to be the next impact time with this object.  Three special
cases for this geometry are supported: $L = 0$ is a ring
(\sect{ring}); $R = 0$ is a line (so there is no inner surface); and
$L = 0 = R$ is a point, which is treated as a special case of an
infinitesimal ring.

Collision resolution is the same as for the infinite cylinder, unless
the particle has made contact with either end, in which case the
collision is treated as a ring bounce (\sect{ring}).

\subsubsection{Ring} \label{s:ring}

This is a special degenerate case of a zero-length cylinder, but it is
also needed for both the perimeter of a finite disk and the ends of a
finite cylinder.  The parameters are inherited from whatever base type
needs to consider ring collisions, \ie in addition to the origin
$\bvec{O}$ and orientation $\bhat{N}$, the parameters are the radius
$R$ of either the finite disk or the finite cylinder, and either
$\bvec{V}$, $A$, and $\Omega$ for the disk or $\Omega$ for the
cylinder---see the corresponding sections above.

Solving for the impact time of a sphere of radius $s$ with a thin ring
is equivalent to finding the first intersection of a ray with a torus
of tube radius $s$, a common problem in computer ray tracing.  We use
the approach developed by Wagner (2004), for which the possible impact
times $t$ are roots of the quartic equation
\begin{equation}
  a_4 t^4 + a_3 t^3 + a_2 t^2 + a_1 t + a_0 = 0 ,
\end{equation}
where
\begin{eqnarray}
  a_4 & = & \alpha^2 \\
  a_3 & = & 2 \alpha \beta \\
  a_2 & = & \beta^2 + 2 \alpha \gamma + 4 R^2 \nu_N^2 \\
  a_1 & = & 2 \beta \gamma + 8 R^2 \rho_N \nu_N \\
  a_0 & = & \gamma^2 + 4 R^2 (\rho_N^2 - s^2) ,
\end{eqnarray}
and
\begin{eqnarray}
  \alpha & = & \bsym{\nu} \vdot \bsym{\nu} \\
  \beta & = & 2 (\bsym{\rho} \vdot \bsym{\nu}) \\
  \gamma & = & \bsym{\rho}\, \vdot \bsym{\rho} - s^2 - R^2 ,
\end{eqnarray}
for which $\bsym{\rho}$ and $\bsym{\nu}$ have the usual definitions
(so the ray is given by $\bsym{\rho} + \bsym{\nu} t$).  However, this
derivation only applies to the special case $\bvec{O} = (0,0,0)$ and
$\bhat{N} = (0,0,1)$, \ie a ring centered at the origin of a Cartesian
coordinate system and lying in the $xy$ plane (so the normal is in the
$+z$ direction).  For the general case, $\bsym{\rho}$ and $\bsym{\nu}$
must first be transformed (rotated) into the torus frame according to
\begin{eqnarray}
  \bsym{\rho}^\prime & = & \bmat{M} \bsym{\rho} \\
  \bsym{\nu}^\prime & = & \bmat{M} \bsym{\nu} ,
\end{eqnarray}
where the primes indicate post-rotation quantities,
\begin{equation}
  \bmat{M} = \trans{\left(\bhat{X}|\bhat{Y}|\bhat{N}\right)} ,
\end{equation}
and $\bhat{X}$ and $\bhat{Y}$ are constructed from $\bhat{N}$ using
Gram-Schmidt orthonormalization (note rotation around the torus
symmetry axis is unimportant, so we are free to choose any $\bhat{X}$
and $\bhat{Y}$ so long as $\bhat{X}$, $\bhat{Y}$, and $\bhat{N}$ are
mutually orthogonal).  Note $\bmat{M} \bhat{N} = (0,0,1)$, as
required.

We solve the quartic using Laguerre's method (Press \etal\ 2007), a
root-finding scheme optimized for polynomials that does not require
prior bracketing of the roots.

There are several special cases that are checked for before committing
to solving the quartic.  In the case of $R = 0$ (a point),
\eqn{twoparticles} applies immediately, with $s_1 = 0$ and $s_2 = s$.
In the case of $\rho_T = 0$ (particle center on the ring symmetry
axis) and $|\nu_N| = |\bsym{\nu}|$ (particle moving parallel to the
symmetry axis), an arbitrary point is chosen on the ring and
\eqn{twoparticles} is used to find the impact time.  Also, as usual,
if $\nu = 0$ ($\alpha = 0$), there can be no collision.

To resolve the collision, \eqntwo{wallbounce1}{wallbounce2} apply, with
\begin{equation}
  \bhat{n} = \left(\bsym{\rho}_{\mathrm{impact}} - R
  \bhat{T}_{\mathrm{impact}}\right)/s . \label{e:ringimpactnormal}
\end{equation}
(So in the degenerate case of $R = 0$, \ie contact with a point,
$\bhat{n} = \bsym{\rho}_{\mathrm{impact}}/s$.)  For the special case
of the particle center lying on the orientation axis, with $R \le s$
(indicating the particle has struck an entire end perimeter perfectly
and thus $\bhat{T}_{\mathrm{impact}}$ is undefined), $\bhat{n} =
\mathrm{sign}\left[\rho_{N,\mathrm{impact}}\right] \bhat{N}$, \ie the
same as for a plane bounce (\sect{infiniteplane}).

\subsection{Overlap Handling} \label{s:overlap}

The derivations above make the tacit assumption that particles are
neither touching nor intersecting any geometric shapes (including
other particles) at the start of the drift interval, \ie legitimate
impact times $t$ should always be positive.  Unfortunately, due to
finite computer precision, sometimes a particle ends up in contact
with or interpenetrating another object at the start of the step, \eg
a collision during the previous step may have been missed, or
round-off error led to an overlap.  The simplest fix is to allow $t$
to be negative (or zero) to correct such situations.  It does,
however, greatly complicate the decision logic when computing $t$ for
the various wall geometries discussed above.

As an example, in the case of the infinite plane, a particle is
touching or overlapping if $\rho_N \le s$, in which case the smallest
non-positive $t$ for impact is chosen.  In practice, a modified
version of \eqn{planeimpact} is used when computing $t$: both faces
are considered regardless of which side the particle center is on,
allowing for the possibility of initial touching/overlap (which is
revealed by one---but not both---of the $t$'s being negative, or one
being negative and the other being zero).  Note that it is still the
case that collisions are ignored if $\nu_N = 0$ or $\rho_N
\nu_N \ge 0$, regardless of whether the particle is
touching or overlapping the surface---generally overlaps are a tiny
fraction of the particle radius, whereas the condition $\rho_N
\nu_N > 0$ with $\rho_N < s$ implies the particle has
penetrated more than a full radius.

In the interest of brevity, we do not provide exhaustive case-by-case
touching/overlap remedies for all geometries here.  It is sufficient
to note that overlaps do happen, particularly in close-packed
situations where a great many collisions can occur during a single
timestep, but the errors are typically very small (many orders of
magnitude smaller than the particle radius) and can be corrected by
allowing $t$ to be negative (or zero).  The test suite described in
\sect{tests} was designed to stress the code and shows that any
overlap artifacts encountered are negligible.  \code{Pkdgrav}
incorporates other overlap strategies for situations involving
interparticle gravity and/or more complex aggregated particle shapes,
but discussion of these strategies is beyond the scope of this paper.

\section{BASIC TESTS} \label{s:tests}

A suite of simple tests was developed during the coding stage to
exercise each new feature/geometry as it was added.  Examples include
dropping particles into a cylinder bisected by an inclined plane,
``rolling'' a perfectly balanced sphere on the inside edge of a
vertical ring, and bouncing hundreds of spheres on an oscillating disk
inside a cylinder.  Here we report on two tests designed to
demonstrate correct dynamic behavior of the granular assembly.  In the
companion paper (Murdoch \etal\ 2011, in preparation), we present a
detailed study of grain motions on a vibrating plate.

\subsection{Model Atmosphere} \label{s:atmos}


For this test, a ball of approximately 1,000 close-packed particles is
dropped from rest into the top half of an infinitely long vertical
cylinder bisected by an infinite horizontal plane.  A uniform gravity
field points vertically downward (there is no interparticle gravity)
and collisions among the particles and between particles and the walls
are elastic (so there is no dissipation).  The particles are roughly
equally divided among three different masses: 1, 3, and 10 (arbitrary
units; for this simulation, the cylinder radius is 1.0, the particle
radius is 0.022 for all masses, the drop height of the center of the
particle ball is 1.0, and the acceleration in the vertical $z$
direction is $-2.5\times 10^9$; internally, the universal constant of
gravitation $G \equiv 1$).  The expected behavior, after the transient
splash, is that the particles achieve energy equipartition, with the
smallest-mass particles reaching a scale height 3 times higher than
that of the medium-mass particles and about 10 times higher than that
of the most massive particles.

\Fig{atmossnaps.pptx} shows ray-traced snapshots of the system, from
the initial condition to the final equilibrium state.  \Fig{atmosevol}
shows the evolution of the mean heights of the three particle mass
populations.  The ``height'' of a particle is the distance separating
its bottom surface from the plane, \ie the $z$ coordinate of its
center minus its radius, with the horizontal plane located at $z = 0$.
Our hypothesis is that the height $z$ of a particle with mass $m$ is
drawn from a probability distribution $P_m(z) = e^{-z/h_m}/h_m$,
where $h_m$ is the scale height associated with particles of mass $m$.
The best estimate of $h_m$
for our model (for a given $m$) is found by maximizing the likelihood
\begin{equation}
  \ln {\cal L} = \sum_{i \in m} \ln P_m(z_i) + \mathrm{const} .
\end{equation}
Substituting for our model $P_m$, we find ${\cal L}$ is maximized when
$h_m = \sum_{i \in m} z_i/N_m$, where $N_m$ is the number of particles
of mass $m$; in other words, the best estimate of $h_m$ is the average
height of all particles with that mass.  This is what is shown in
\fig{atmosevol} (solid lines).  The measured scale heights show the
expected inverse proportionality with mass, but exhibit some
fluctuations.  In fact, a perfect match to the model is not expected
due to finite-size effects: in the limit of large particle sizes
compared to the cylinder diameter, the bottom layers may become fully
occupied, which is not predicted by the functional form of $P_m$ used
here.  To test this, a few runs varying the cylinder-to-particle
diameter ratio were carried out and showed larger discrepancies from
the ideal scale height ratios for smaller diameter ratios.

\begin{center}
  [FIGURE \ref{f:atmossnaps.pptx} GOES HERE]
\end{center}

\begin{center}
  [FIGURE \ref{f:atmosevol} GOES HERE]
\end{center}

We also performed Kolmogorov-Smirnov (K-S) tests to compare the height
data with the expected distributions $P_m(z)$.  \Fig{kstest} shows the
K-S curves (solid lines) for the final snapshot in
\fig{atmossnaps.pptx} compared to the normalized cumulative
probability distributions $C_m(z)/C_m(z_{\mathrm{max}})$ (dotted
lines), where $C_m(z) = \int_0^{z^\prime} P_m(z^\prime)\,dz^\prime = 1
- e^{-z/h_m}$ and $z_{\mathrm{max}}$ is the largest value of $z$
(among particles of mass $m$) for this snapshot.  The normalization
ensures the cumulative probability distribution has a value of unity
at $z = z_{\mathrm{max}}$ (each curve has a different
$z_{\mathrm{max}}$, but the K-S statistic is insensitive to scaling of
the independent variable, which is why there are no tick marks on the
horizontal axis of \fig{kstest}).  For this particular snapshot, which
represents an arbitrary instant after equilibrium has been reached, we
find K-S probabilities of 0.89, 0.08, and 0.70, corresponding to
masses 1, 3, and 10, meaning we cannot reject the null hypothesis that
the data is drawn from $P_m$ at those confidence levels.  For the
middle mass, this is not very convincing, but we find there is
significant variation between snapshots.  Taking the average
probabilities from 50\%, 60\%, 70\%, 80\%, 90\%, and 100\% of the way
through the simulation, we find mean K-S probabilities of $0.77 \pm
0.27$, $0.40 \pm 0.33$, and $0.72 \pm 0.16$ (the errorbars denote 1
standard deviation from the mean).  Moreover, if we repeat the
analysis for a cylinder that is half as wide, the probabilities drop
to $0.22 \pm 0.31$, $0.05 \pm 0.06$, and $0.62 \pm 0.35$.  We must
caution that because the mean and maximum $z$ values measured from the
data were used in the model for the K-S test, the distribution of the
K-S statistic may not be fully consistent with the formula used to
compute the null hypothesis probability (\cf Press \etal\ 2007).
Repeating the analysis using the \emph{expected} values of $h_m$ (see
below), the mean K-S probabilities for the radius 1.0 cylinder do not
change significantly.  Regardless, there is no way to avoid using
$z_{\mathrm{max}}$ to insure the cumulative probability is normalized,
so the K-S test probabilities must be considered formal values only.

\begin{center}
  [FIGURE \ref{f:kstest} GOES HERE]
\end{center}

Finally, we can apply the principles of hydrostatic equilibrium and
the energy equipartition theorem to derive the expected scale heights
$h_m$ and compare these with the measured values for each mass
population.  Treating the particles as an ideal gas at constant
temperature $T$ (an ideal isothermal atmosphere), the number density
$n$ of particles is proportional to the pressure $P$.  In hydrostatic
equilibrium,
\begin{equation}
  dP = - \rho g \, dz ,
\end{equation}
where $\rho = mn$ and $g$ is the (assumed constant) gravitational
acceleration.  Substituting $\rho = m/(k_BT)$ from the ideal gas law,
where $k_B$ is the Boltzmann constant, and integrating, we find
\begin{equation}
  P = P_0 \exp \left( - \frac{mg}{k_BT} z \right) ,
\end{equation}
which, since $P$ is proportional to $\rho$, gives the desired scale
height
\begin{equation}
  h_m = \frac{k_BT}{mg} . \label{e:scaleheight}
\end{equation}
Now, the total energy of the system, which remained constant to better
than 1 part in $10^4$ for the duration of all the model atmosphere
tests we performed (the starting value was $2.9869 \times 10^{15}$ in
system units), should be roughly equally divided among the particles
once equilibrium is achieved.  The energy of a given particle in our
system is
\begin{equation}
  E = \frac{1}{2} m v^2 + mgz ,
\end{equation}
where $v$ is the particle speed.  Applying the equipartition theorem,
which states in part that for each degree of freedom $x$ that
contributes $x^a$ to the energy, the average energy corresponding to
that degree of freedom at thermal equilibrium is $k_BT/a$, the average
energy per particle is
\begin{equation}
  <E>\, = \frac{3}{2} k_BT + k_BT = \frac{5}{2} k_BT
\end{equation}
(recall $v$ incorporates three degrees of freedom, one for each
component of the velocity vector in three dimensions).  Equating this
to the total energy divided by the number of particles (exactly 991 in
our simulation), solving for $k_BT$ and substituting into
\eqn{scaleheight}, we get expected $h_m$ values of 1.88, 0.626, and
0.188 for our particle masses of 1, 3, and 10, respectively (notice
these are in the correct 10:3:1 ratio; the values are shown as the
dotted lines in \fig{atmosevol}).  Averaging as we did previously for
the K-S probabilities, we find measured $h_m$ values of $1.89 \pm
0.06$, $0.628 \pm 0.045$, and $0.185 \pm 0.007$ (\cf \fig{atmosevol}),
in quite good agreement.  For the simulation where the radius is
halved, the values are $1.89 \pm 0.15$, $0.720 \pm 0.034$, and $0.204
\pm 0.011$, in worse agreement, as might be expected due to the
greater severity of the finite-size effect.

In summary, we find that this test of the integrator with particle and
wall bouncing gives results consistent with what would be expected in
a real system.

\subsection{Tumbler} \label{s:tumbler}

Brucks \etal\ (2007) carried out a series of laboratory experiments to
measure the dynamic angle of repose $\beta$ of glass beads in a
rotating drum (tumbler) at various effective gravitational
accelerations.  The beads had \emph{diameter} (not radius) $d = 0.53
\pm 0.05$ mm and occupied about 50\% of the volume of a tumbler of
radius $R$ = 30 mm and length $L$ = 5 mm.  A tumbler with $R$ = 45 mm
was also used, but we restrict our comparison to the smaller one for
this test.  The inner wall of the cylinder was lined with sandpaper to
provide a rough surface.  A centrifuge was used to provide effective
gravitational accelerations between 1 and 25 $g$ (where $g$ = 9.81 m
s$^{-2}$ is Earth-norm gravity).  The angular rotation speed of the
tumbler $\Omega$ was varied to achieve a range of Froude numbers
(``Fr'') for the system, where Fr is defined as the ratio of
centripetal to effective gravitational acceleration at the cylinder
periphery:\footnote{Occasionally the square root of the quantity on
  the right-hand-side of \eqn{Froude} is found in the literature, \ie
  $\mbox{Fr} = \Omega \sqrt{R/g}$; we use the expression in
  \eqn{Froude} to compare our results more easily with Brucks
  \etal\ (2007).}
\begin{equation}
  \mbox{Fr} \equiv \frac{\Omega^2R}{g_{\mathrm{eff}}} . \label{e:Froude}
\end{equation}

Figure 3 of Brucks \etal\ (2007) shows $\beta$ as a function of Fr for
various $g_{\mathrm{eff}}$ obtained in their experiment.  Our aim for
this test was to compare simulated results for $g_{\mathrm{eff}}/g =
1$ and $\mbox{Fr} \ge 0.001$.  Smaller Fr numbers are computationally
challenging due to the limited motion of the particles (but the
$\beta$ curve is essentially flat for Fr $\lesssim$ 0.01 anyway).
Different $g_{\mathrm{eff}}$, particularly $g_{\mathrm{eff}}/g < 1$ (a
regime Brucks \etal\ 2007 could not explore), will be the subject of
future work.


We constructed the simulated experimental apparatus using an infinite
horizontal cylinder of radius $R$ (30 mm) cut into a short segment by
two infinite vertical planes separated by a distance $L$ (5 mm).  For
the purpose of this test, the planes were ``frictionless''
($\varepsilon_n = \varepsilon_t = 1$).  To keep the tests tractable, a
fixed particle \emph{radius} of $s = 0.53$ mm was used, \ie twice the
size of the glass beads used in the experiment (thereby providing
nearly a factor of 10 savings in particle number, so each run would
take of order a day, instead of more than a week).  The initial
conditions were prepared in stages, alternately filling the ``bottom''
of the cylinder (\ie by removing one plane and having gravity point
down the length of the cylinder), closing and rotating the cylinder to
allow particles to stick uniformly to the inner cylinder wall (to
mimic the roughness provided by the sandpaper in the real experiment),
refilling to get the desired volume fraction (about 50\%), rotating
some more, and then allowing the particles to settle.  There were 6513
particles total, of which 568 ended up stuck to the cylinder
wall.\footnote{In order to allow for the finite diameter of the stuck
  particles, the cylinder was actually $R + 2s$ = 31.06 mm in radius.}
This starting condition was used for all the runs.  For particle
collisions, the normal coefficient of restitution $\varepsilon_n =
0.64$ (an estimate for glass beads derived by Hei{\ss}elmann
\etal\ 2009) and the tangential coefficient of restitution
$\varepsilon_t = 0.8$ (to provide some roughness; this value is poorly
constrained).

As an aside, initially this experiment was conducted with surface
coupling (\ie $\varepsilon_t < 1$ for the cylinder) instead of
sticky walls.  Although, for sufficiently fast cylinder spin, the
particles also start to tumble in this setup, the particle spins
quickly align with the cylinder spin (so that $\omega_i s \sim \Omega
R$, within a factor of a few) and the particles settle to the bottom
of the cylinder, sloshing around slightly.  A rough surface instead
provides a constant, somewhat random perturbation that ensures the
tumbling behavior persists among the loose particles.

\Fig{tumbsnaps.pptx} shows ray-traced snapshots after achieving
equilibrium for 8 different Froude numbers (0.001, 0.01, 0.05, 0.1,
0.3, 0.5, 1.0, and 1.5).  For the sub-critical cases ($\mbox{Fr} <
1$), the total simulated run time was about 2.5 s, with a fixed
timestep of about 5 $\mu$s.  For Fr = 1.0 and 1.5, the run time was
$\sim$0.5 s with a timestep of $\sim$1 $\mu$s (these latter cases
achieve equilibrium faster, but the more rapid cylinder rotation
dictated a smaller timestep).  Also shown for the sub-critical cases
is the estimated slope, which was measured by fitting straight lines
to sample surface points over several snapshots and taking the
average.  Only particles to the left of center were considered for
these measurements, as particles tend to pile up on the right.  As
expected, for Fr = 1.0, it can be seen that centrifuging has begun; by
Fr = 1.5, a thin layer of free particles persists along the inner
surface of the upper-right quadrant of the cylinder.  For $\mbox{Fr} <
1$, $\beta$ depends on the rotation rate, with roughly the expected
dependence from the experiments.  A study by Taberlet \etal\ (2006)
shows that the shape of the granular pile in these kinds of
experiments depends on interactions with the end caps, with friction
leading to ``S'' shapes and lack of friction leading to straighter
slopes (as in our case).  They also found that larger $L$ (\ie using a
longer cylinder, for a fixed particle size) leads to straighter
slopes.  Testing such dependencies in our approach will be the subject
of future work.

\begin{center}
  [FIGURE \ref{f:tumbsnaps.pptx} GOES HERE]
\end{center}

\Fig{tumbplot} shows the measured values of $\beta$ from the
simulations (filled squares), with 1-$\sigma$ uncertainty errorbars,
along with data taken from Fig.\ 3 of Brucks \etal\ (2007) for the $R$
= 30 mm tumbler.  The experimental data show systematically larger
dynamic slopes compared with the simulated data, but both follow a
similar trend with Froude number.  Some of the discrepancy in slope
may arise from the different $R/s$ ratio for the data sets, namely
about 110 for the experiment while only 57 for the simulations.
Increasing the ratio for the simulations to match the experiments
would require a factor of 8 more free particles (as well 4 times more
stuck particles), which was beyond the scope of these tests.  We note
that the $R$ = 45 mm tumbler used in the experiments, with $R/s \sim
170$, exhibits equilibrium dynamic slopes about 10 deg higher than for
$R$ = 30 mm (Brucks \etal\ 2007, Fig.\ 3).

\begin{center}
  [FIGURE \ref{f:tumbplot} GOES HERE]
\end{center}

There are other caveats to consider as well.  The simulated particles
are identical in size and are perfectly spherical; deviations from
uniformity of the glass beads used in the experiments may increase the
dynamic slopes (this could also give rise to a higher slope at small
Froude numbers, where the experimental results did not appear to be as
sensitive to $R/s$).  Furthermore, the adopted $\varepsilon_n$ and
$\varepsilon_t$ values for the simulations are mostly educated
guesses; more experimentation is needed.  The simulation used
frictionless confining planes, whereas the flat walls in the
experiment were comprised of a metal plate and a glass plate, both
constrained to rotate with the cylinder, and neither of which were
frictionless.  In the simulations, the cylinder rotation rate was
imposed instantaneously at the start, which may have caused spurious
bulking as the assemblage reacted to the sudden shear stress.  A
better approach (not implemented here) would be to accelerate the
tumbler gradually to the desired rotation rate.  A related issue is
that the dynamic angle of repose may be affected by the initial
packing fraction of the particles, which depends on the dominant
interparticle forces.  In the absence of interparticle friction,
random close packing is expected, whereas with friction, random loose
packing is a more likely starting condition (Makse \etal\ 2000), and
may lead to a different dynamic angle of repose.  This difference was
not explored here but will be part of a future investigation.
Finally, certain minor run parameters required to keep the simulations
tractable in the particle-based, collision-prediction approach may
result in excessive particle energy, leading to shallower simulated
slopes; experimentation with these thresholds is needed.

\section{DISCUSSION} \label{s:disc}

The model atmosphere scenario presented in \sect{atmos} is an example
of a dilute granular flow in which particle-particle and particle-wall
interactions are dominated by near-instantaneous, inelastic, two-body
collisions.  Our numerical approach, in the context of granular
dynamics literature, can be characterized as hard-sphere DEM (see
\sect{intro}), which is particularly well-suited to simulating dilute
granular flows (Mehta 2007).  In the dilute regime, soft- and
hard-particle methods should give equivalent results, but so-called
event-driven simulations, in which no computations are needed between
collision events, are typically much faster (Luding 2004).  However,
within dense granular flows in which many-body interactions dominate
and particles have long-lived contacts with many neighbors, it is
possible that the assumptions of the event-driven and hard-sphere DEM
models fail (Delannay \etal\ 2007).

In a rotating tumbler (\cf \sect{tumbler}), there exists a region of
long-lived contacts but also a faster flowing layer.  By varying the
Froude number, it is possible to investigate many different regimes,
from the fast-flowing regime with a large Froude number to the regime
when the Froude number is low and most contacts are longer-lived.
Thus the tumbler is an intermediate regime between dilute and dense
flows, providing a convenient test of the range of applicability of
our hard-sphere method.  Indeed, the tumbler has been studied in
detail with many different numerical techniques; we provide a brief
overview in the following.

McCarthy \etal\ (1996, 2000) and McCarthy and Ottino (1998) used a
soft-sphere model to compare simulations of non-cohesive granular
materials in a slowly rotating drum to experimental data.  In order to
limit the number of particles necessary in the simulations they
proposed a ``hybrid'' simulation technique.  This method involves
using a geometrical model to determine the bulk motion of the
particles and then performing particle dynamics simulations on only
the particles contained within and bordering the avalanching wedge.
They found a favorable match between experiment and simulation, at low
computational cost, but with the tradeoff that the method is tailored
to a very specific geometry.

Khakhar \etal\ (1997) used a similar technique to separate the
granular material in the rotating drum into a ``rapid flow region''
and a ``fixed bed.''  A continuum model in which averages are taken
across the layer was used to analyze behavior of the flowing layer.
The motion of grains on the free surface of a granular mixture in a
rotating drum was also investigated by Monetti \etal\ (2001) using
Monte Carlo simulations of a two-dimensional lattice gas model.  The
model takes into account rotational frequency, frictional forces, and
the gravitational field.

Soft-sphere methods have also repeatedly and successfully been used to
describe granular material in a rotating drum (Dury \etal\ 1998; Dury
and Ristow 1999; Rapaport 2002; Pohlman \etal\ 2006; Taberlet
\etal\ 2006).  This has also extended been to two-dimensional
simulations of irregular-shaped particles by Poschel and Buchholtz
(1995).

Hard-sphere methods have not been commonly used for rotating drums,
however Gui and Fan (2009) did perform numerical simulations of motion
of rigid spherical particles within a 2-D tumbler with an inner
wavelike surface.  The rotation of the tumbler was simulated as a
traveling sine wave around a circle.

The observed behavior in our tumbler simulations shows that the code
correctly models the transitions from global regimes with increasing
Froude number.  The differences between experiment and simulation are
larger at lower Froude number and start to decrease at higher Froude
number.  This suggests that our hard-sphere model is more suited to
the dilute flow regimes of high Froude number rather than dense flow
regimes of low Froude number.  Soft-sphere DEM is typically used for
dense granular flows (Zhu \etal\ 2007), as the detailed
characterization of interparticle contacts is more suitable for this
regime.  However, hard-sphere methods can approach experimental and
soft-sphere simulation results for very low friction materials.  For
example, Pohlman \etal\ (2006) found an angle of repose of only
16.9$^\circ$, for low friction ($\mu = 0.1$) chrome steel beads at a
Froude number of $4.94 \times 10^{-5}$.  This is within a few degrees
of our expected results at this Froude number.

In summary, our approach currently favors dilute or fluid flows over
dense granular flows, but we are in the process of implementing a
soft-sphere method.  Whether regolith in low-gravity environments,
such as the surface of an asteroid, when reacting to external
stresses, is more appropriately modeled as a dilute or dense granular
flow is the subject of ongoing investigation.  Regardless, numerical
methods will play an important role in the study of these
environments.

\section{CONCLUSIONS} \label{s:concl}

We have added new features to an existing, well-tested $N$-body code
that allow convenient modeling of granular dynamics in a variety of
conditions.  Details of collision detection and resolution involving
four principal supported wall geometries (infinite plane, finite disk,
infinite cylinder, and finite cylinder) and associated degenerate
cases (point, line, and ring) were provided.  Two dynamical test
suites were presented, one for a model ``atmosphere'' showing the
correct equilibrium distribution of scale heights among a polydisperse
population of dissipationless balls, and the other for a tumbler that
showed correct qualititative behavior as a function of surface wall
properties and cylinder rotation rate.  Application of these new
capabilities to numerical simulations designed to investigate
collective grain behavior on vibrating plates will be presented in
Murdoch \etal\ (2011, in preparation).  Planned future code
development includes adding support for rigid and semi-rigid particle
aggregates (\cf Richardson \etal\ 2009) that would allow study of more
complex particle shapes in simulations of granular dynamics, and
implementing a soft-sphere model to better match the properties of
dense granular flows.

\newpage
\section*{Acknowledgments}

This material is based upon work supported by the National Aeronautics
and Space Administration under Grant No.\ NNX08AM39G issued through
the Office of Space Science and by the National Science Foundation
under Grant No.\ AST0524875.  KJW was supported by a Poincar\'e
Fellowship at Observatoire de la C\^ote d'Azur (OCA).  PM and NM
acknowledge financial support from the French Programme National de
Plan\'etologie.  NM acknowledges financial support from Thales Alenia
Space and The Open University.  DCR thanks M. C. Miller for helpful
discussions.  We thank Dan Scheeres and an anonymous reviewer for their comments and suggested improvements.  Some simulations in this work were performed on the
CRIMSON Beowulf cluster at OCA.  Raytracing for
\figtwo{atmossnaps.pptx}{tumbsnaps.pptx} was carried out using the
Persistence of Vision
Raytracer.\footnote{\code{http://www.povray.org/}} Data points from
Fig.\ 3 of Brucks \etal\ (2007) were extracted using Plot
Digitizer.\footnote{\code{http://plotdigitizer.sourceforge.net/}}

\newpage
\section*{References}

\begin{description}

\item \paper {Asphaug, E., King, P.J., Swift, M.R., Merrifield, M.R.}
  2001 {Brazil nuts on Eros: Size-sorting of asteroid regolith}
  {Proc.\ Lunar Sci.\ Conf.} 32 1708

\item \paper {Brucks, A., Arndt, T., Ottino, J., Lueptow, R.} 2007
  {Behavior of flowing granular materials under variable $g$}
  {Phys.\ Rev. E} 75 032301-1--032301-4

\item \paper {Cleary, P.W., Sawley, M.L.} 2002 {DEM modelling of
  industrial granular flows: 3D case studies and the effect of
  particle shape on hopper discharge} {Applied Mathematical Modelling}
  26 89--111

\item \paper {Delannay, R., Louge, M., Richard, P., Taberlet, N.,
  Valance, A.} 2007 {Towards a theoretical picture of dense granular
  flows down inclines} {Nature Materials} 6 99--108

\item \paper {Dury, C.M., Ristow, G.H.} 1999 {Competition of mixing
  and segregation in rotating cylinders} {Phys.\ Fluids} 11 1387--1394

\item \paper {Dury, C.M., Ristow, G.H., Moss, J.L., Nakagawa, M.}
  1998 {Boundary effects on the angle of repose in rotating cylinders}
  {Phys.\ Rev.\ E} 57 4491--4497

\item \paper {Elaskar, S.A., Godoy, L.A., Gray, D.D., Stiles, J.M.}
  2000 {A viscoplastic approach to model the flow of granular solids}
  {International Journal of Solids and Structures} 37 2185--2214

\item Falk, M.L., Toiya, M., Losert, W., 2008. Shear transformation
  zone analysis of shear reversal during granular flow. ArXiv e-prints
  0802.1752.

\item \paper {Fraige, F.Y., Langston, P.A., Chen, G.Z.} 2008 {Distinct
  element modelling of cubic particle packing and flow} {Powder
  Technology} 186 224--240

\item \paper {Goddard, J.D.} 1990 {Nonlinear elasticity and
  pressure-dependent wave speeds in granular media} {Proc.\ Royal
  Soc.\ London} A430 105--131

\item \paper {Gui, N., Fan, J.} 2009 {Numerical simulation of motion
  of rigid spherical particles in a rotating tumbler with an inner
  wavelike surface} {Powder Technology} 192 234--241

\item \paper {Hei{\ss}elmann, D., Blum, J., Fraser, H., Wolling, K.}
  2009 {Microgravity experiments on the collisional behavior of
    saturnian ring particles} Icarus 206 424--430

\item \paper {Holsapple, K.A.} 2001 {Equilibrium configurations of
  solid cohesionless bodies} Icarus 154 432--448

\item \paper {Holsapple, K.A.} 2004 {Equilibrium figures of spinning
  bodies with self-gravity} Icarus 172 272--303

\item \paper {Holsapple, K.A., Michel, P.} 2006 {Tidal disruptions: A
  continuum theory for solid bodies} Icarus 183 331--348

\item \paper {Holsapple, K.A., Michel. P.} 2008 {Tidal
  disruptions. II. A continuum theory for solid bodies with strength,
  with applications to the solar system} Icarus 193 283--301

\item \paper {Hong, D.C., McLennan, J.A.} 1992 {Molecular dynamics
  simulations of hard sphere granular particles} {Physica A:
  Statistical Mechanics and its Applications} 187 159--171

\item \paper {Huilin, L., Yunhua, Z., Ding, J., Gidaspow, D., Wei, L.}
  2007 {Investigation of mixing/segregation of mixture particles in
    gas-solid fluidized beds} {Chem.\ Engin.\ Sci.} 62 301--317

\item \paper {Kacianauskas, R., Maknickas, A., Kaceniauskas, A.,
  Markauskas, D., Balevicius, R.} 2010 {Parallel discrete element
  simulation of poly-dispersed granular material} {Advances in
  Engineering Software} 41 52--63

\item \paper {Kamrin, K., Bazant, M.Z.} 2007 {Stochastic flow rule for
  granular materials} {Phys.\ Rev.\ E} 75 041301

\item \paper {Kharhar, D.V., McCarthy, J.J., Shinbrot, T., Ottino,
  J.M.} 1997 {Transverse flow and mixing of granular material in a
  rotating cylinder} {Phys. Fluids} 9 31--43

\item \paper {Kosinski, P., Hoffmann, A.C.} 2009 {Extension of the
  hard-sphere particle-wall collision model to account for particle
  deposition} {Phys.\ Rev.\ E} 79 061302

\item \paper {Latham, J-P., Munjiza, A., Garcia, X., Xiang, J.,
  Guises, R.} 2007 {Three-dimensional particle shape acquisition and
  use of shape library for DEM and FEM/DEM simulation} {Minerals
  Engineering} 21 797--805

\item \paper {Losert, W., Bocquet, L., Lubensky, T. C., Gollub, J.P.}
  2000a {Particle dynamics in sheared granular matter}
  {Phys.\ Rev.\ Lett.} 85 1428--1431

\item \paper {Losert, W., G\'eminard, J.C., Nasuno, S., Gollub, J.P.}
  2000b {Mechanisms for slow strengthening in granular materials}
  {Phys.\ Rev.\ E} 61 4060--4068

\def\chap#1 #2 #3 #4 #5 #6 #7 {#1, #2. #3. In: #4 (Eds.), #5. #6, pp.\ #7.}

\item \chap {Luding, S.} 2004 {Molecular dynamics simulations of
  granular materials} {Hinrichsen, H., Wolf, D.E.} {The Physics of
  Granular Media} {Wiley-VCH, Weinheim} 299--323 REVISE CHECK END PAGE

\item \paper {Makse, H.A., Johnson, D.L., Schwartz, L.M.} 2000
  {Packing of compressible granular materials} {Phys.\ Rev.\ Let.} 84
  4160--4163

\item \paper {McCarthy, J.J., Ottino, J.M.} 1998 {Particle dynamics
  simulation: A hybrid technique applied to granular mixing} {Powder
  Technology} 97 91--99

\item \paper {McCarthy, J.J., Shinbrot, T., Metcalfe, G., Wolf, E.J.,
  Ottino, J.M.} 1996 {Mixing of granular materials in slowly rotated
  containers} {AIChE Journal} 42 3351--3363

\item \paper {McCarthy, J.J., Khakhar, D.V., Ottino, J.M.}  2000
  {Computational studies of granular mixing} {Powder Technology} 109
  72--82



\item \book {Mehta, A.J.} 2007 {Granular Physics} {Cambridge
  Univ.\ Press} {New York}

\item \paper {Michel, P., O'Brien, D., Abe, S., Hirata, N.} 2009
  {Itokawa's cratering record as observed by Hayabusa: Implications
    for its age and collisional history} Icarus 200 503--513

\item Miwa, Y., Yano, H., Morimoto, M., Mori, O., Kawaguchi, J.,
  2008. A study on constitution of asteroids with Brazil-nut
  effect. Proc.\ 26th International Symposium on Space Technology and
  Science 2008, 2008-k-03.

\item \paper {Miyamoto, H. and 14 colleagues} 2007 {Regolith migration
  and sorting on asteroid Itokawa} Science 316 1011--1014

\item \paper {Monetti, R., Hurd, A., Kenkre, V.M.} 2001 {Simulations
  for dynamics of granular mixtures in a rotating drum} {Granular
  Matter} 3 113--116

\item \paper {Mueth, D.M., Debregeas, G.F., Karczmar, G.S., Eng, P.J.,
  Nagel, S.R., Jaeger, H.M.} 2000 {Signatures of granular
  microstructure in dense shear flows} Nature 406 385--389

\item \inprep {Murdoch, N., Berardi, C., Michel, P., Richardson, D.C.,
  Wolfgang, L., Green, S.F.} 2011 {Numerical simulations of granular
  material dynamics. II. Comparison with shaking experiments} Icarus

\item \book {Press, W.H., Teukolsky, S.A., Vetterling, W.T., Flannery,
  B.P.} 2007 {Numerical Recipes: The Art of Scientific Computing,
  third ed.} {Cambridge Univ.\ Press, New York}

\item \paper {Pohlman, N.A., Severson, B.L., Ottino, J.M., Lueptow,
  R.M.} 2006 {Surface roughness effects in granular matter: Influence
  on angle of repose and the absence of segregation} {Phys.\ Rev.\ E}
  73 031304

\item Poschel, T., Buchholtz, V., 1995. Complex flow of granular
  material in a rotating cylinder. Chaos, Solitons \& Fractals 5,
  1901--1905, 1907-1912. 

\item \paper {Rapaport, D.C.} 2002 {Simulational studies of axial
  granular segregation in a rotating cylinder} {Phys.\ Rev.\ E} 65
  061306

\item \paper {Richardson, D.C.} 1994 {Tree code simulations of
  planetary rings} {Mon.\ Not.\ R. Astron.\ Soc.} 269 493--511

\item \paper {Richardson, D.C., Quinn, T., Stadel, J., Lake, G.} 2000
  {Direct large-scale $N$-body simulations of planetesimal dynamics}
  Icarus 143 45--59

\item \paper {Richardson, D.C., Michel, P., Walsh, K.J., Flynn, K.W.}
  2009 {Numerical simulations of asteroids modeled as gravitational
    aggregates} {Plan.\ \& Space Sci.} 57 183--192

\item \paper {Richardson, J.E., Melosh, H.J., Greenberg, R.} 2004
  {Impact-induced seismic activity on asteroid 433 Eros: A surface
    modification process} Science 306 1526--1529

\item \paper {Robinson, M.S., Thomas, P.C., Veverka, J., Murchie,
  S.L., Wilcox, B.B.} 2002 {The geology of 433 Eros}
  {Meteorit.\ Planet.\ Sci.} 37 1651--1684

\item \paper {Rosato, A., Strandburg, K.J., Prinz, F., Swendsen, R.H.}
  1987 {Why the Brazil nuts are on top: Size segregation of
    particulate matter by shaking} {Phys.\ Rev.\ Lett.} 58 1038--1040

\item S\'anchez, P., Scheeres, D.J., Swift, M.R., 2010. Impact driven
  size sorting in self-gravitating granular aggregates. Lunar
  Planet.\ Sci.\ 41. Abstract 1533. 

\item \paper {Savage, S.B.} 1998 {Analyses of slow high-concentration
  flows of granular materials} {J. Fluid Mech.} 377 1--26

\item \paper {Scheeres, D.J., Abe, M., Yoshikawa, M., Nakamura, R.,
  Gaskell, R.W., Abell, P.A.} 2007 {The effect of YORP on Itokawa}
  Icarus 188 425--429

\item \paper {Scheeres, D.J., Hartzell, C.M., S\'anchez, P., Swift,
  M.} 2010 {Scaling forces to asteroid surfaces: The role of cohesion}
  Icarus 210 968--984

\item \paper {Sharma, I., Jenkins, J.T., Burns, J.A.} 2006 {Tidal
  encounters of ellipsoidal granular asteroids with planets} Icarus
  183 312--330

\item \paper {Sharma, I., Jenkins, J.T., Burns, J.A.} 2009 {Dynamical
  passage to approximate equilibrium shapes for spinning, gravitating
  rubble asteroids} Icarus 200 304--322

\item \thesis {Stadel, J.} 2001 {Cosmological $N$-body simulations and
  their analysis} {University of Washington, Seattle} 126

\item \inpress {Szarf, K., Combe, G., Villard, P.} 2010 {Polygons
  vs.\ clumps of discs: A numerical study of the influence of grain
  shape on the mechanical behaviour of granular materials} {Powder
  Technology} 

\item \paper {Taberlet, N., Richard, P., Hinch, J.E.} 2006 {S shape of
  a granular pile in a rotating drum} {Phys.\ Rev.\ E} 73 050301(R)

\item \paper {Toiya, M., Stambaugh, J., Losert, W.} 2004 {Transient
  and oscillatory granular shear flow} {Phys.\ Rev.\ Lett.} 93 088001

\item \paper {Wada, K., Senshu, H,, and Matsui, T.} 2006 {Numerical
  simulation of impact cratering on granular material} Icarus 180
  528--545

\item Wagner, M. 2004. Ray/torus intersection.  Independent online
  publication.\footnote{\code{http://www.emeyex.com/site/projects/raytorus.pdf}}

\item \paper {Walsh, K.J., Richardson, D.C., Michel, P.} 2008
  {Rotational breakup as the origin of small binary asteroids} Nature
  454 188--19

\item \paper {Zhu, H.P., Zhou, Z.Y., Yang, R.Y., Yu, A.B.} 2007
  {Discrete particle simulation of particulate systems: Theoretical
    developments} {Chem.\ Engin.\ Sci.} 62 3378--3396

\end{description}

\newpage
\section*{Figure Captions}

\begin{description}
  
  \figcap{spheres.pptx} Diagram illustrating the quantities used
  for predicting the collision of two spheres (\cf
  \eqns{sphcolcomp}{twoparticles}).  Although the diagram depicts a
  collision in two dimensions, the derivations apply to the full
  three-dimensional case.

  \figcap{plane.pptx} Geometry of sphere-plane collision.  Here the
  plane is depicted as a finite, thick disk for illustration only; in
  reality the plane has zero thickness and infinite extent.  For
  simplicity, velocity vectors are not shown.  The unit vector
  $\bhat{T}$ is in the plane, while $\bhat{N}$ is perpendicular to it
  (for clarity of presentation, the vector representing $\bhat{T}$ is
  given length $|\rho_T|$, \ie the distance from the plane origin to
  the contact point).  Note $\bhat{n} = - \bhat{N}$ and $\bsym{\rho} =
  \rho_N \bhat{N} + \rho_T \bhat{T}$, where in this scenario $\rho_N$
  must equal $s$, the radius of the impacting sphere.  The geometry is
  similar for a sphere-disk collision (\sect{finitedisk}), where the
  disk in the figure would have radius $R$ but still have no
  thickness.

  \figcap{cylinder.pptx} Geometry of sphere-cylinder collision.  Shown
  is a finite cylinder of radius $R$ and length $L$, with origin
  $\bvec{O}$ at the geometric center.  For an infinite cylinder, $L
  \rightarrow \infty$ and $\bvec{O}$ can be anywhere along the
  symmetry axis.  For ease of illustration, only the case of exterior
  impact is shown; interior impact is similar, except $\bhat{n}$ has
  opposite sign.  The unit vector $\bhat{T}$ is the perpendicular
  pointing from the cylinder axis to the point of contact, while
  $\bhat{N}$ points along the cylinder axis (for clarity of
  presentation, the vectors representing $\bhat{T}$ and $\bhat{n}$ are
  given length $R$ and $s$, respectively, where $s$ is the radius of
  the impacting sphere).  Note $\bhat{n} = \bhat{T}$ (for exterior
  impact) and $\bsym{\rho} = \rho_N \bhat{N} + \rho_T \bhat{T}$, where
  in this scenario $\rho_T$ must equal $R + s$; $\rho_N$ is simply
  $\bsym{\rho} \vdot \bhat{N}$.

  \figcap{atmossnaps.pptx} Snapshots from the model atmosphere test
  (\sect{atmos}) showing (a) the initial condition, (b) splash, and (c) final
  equilibrated state.  The particle colors red, magenta, and blue
  correspond to particle masses 1, 3, and 10, respectively (particle
  radii are the same for all masses).  The scale heights of the 3
  populations are expected to be inversely proportional to their
  masses.  The full vertical extent of the smallest-mass particle
  distribution is not shown.  Gravity is directed down in the figure.

  \figcap{atmosevol} Evolution of the mean particle heights in the
  model atmosphere test.  The colors red (top line), magenta (middle
  line), and blue (bottom line) correspond to particle masses 1, 3,
  and 10, respectively.  The dotted horizontal lines are the
  corresponding expected scale heights assuming hydrostatic
  equilibrium, the ideal gas law, and energy equipartition.  The
  timestep for these simulations was $10^{-8}$ in system units.  The
  snapshots of \fig{atmossnaps.pptx} correspond to timesteps 0, 30,
  and 1,000, respectively.

  \figcap{kstest} Cumulative probability distributions for the height
  data (solid lines) and model (dotted lines) corresponding to the
  final snapshot in \fig{atmossnaps.pptx}.  The colors red (middle
  lines at small $z$), magenta (bottom lines at small $z$), and blue
  (top lines) correspond to particle masses 1, 3, and 10,
  respectively.  The K-S statistic measures the maximum vertical
  deviation of the data from the model (see text).

  \figcap{tumbsnaps.pptx} Snapshots of the simulated tumbler
  experiments taken at the end of each run.  The Froude number
  (\eqn{Froude}) is indicated.  View size is about 7.5 cm across each
  frame.  The view is down the cylinder axis, and rotation is in the
  clockwise rotation direction (at this viewing angle, the short
  cylinder itself cannot be seen).  The blue particles along the inner
  surface of the cylinder are ``glued'' in place to provide roughness.
  Where shown, the yellow line indicates an estimate of the surface
  slope (also see \fig{tumbplot}).

  \figcap{tumbplot} Plot of dynamic repose angle as a function of
  Froude number for the simulated tumblers (filled squares) and from
  the experiments of Brucks \etal\ (2007; open squares connected with
  dashed line, from their Fig.\ 3).  Errorbars for the simulated
  points are 1-$\sigma$ uncertainties from averaging slope estimates
  over several snapshots.  The simulated points are shifted down and
  to the right compared with the experiments, which may be due partly
  to the difference in particle-to-tumbler radius ratio.  See main
  text for discussion.

\end{description}

%
%

\putfig{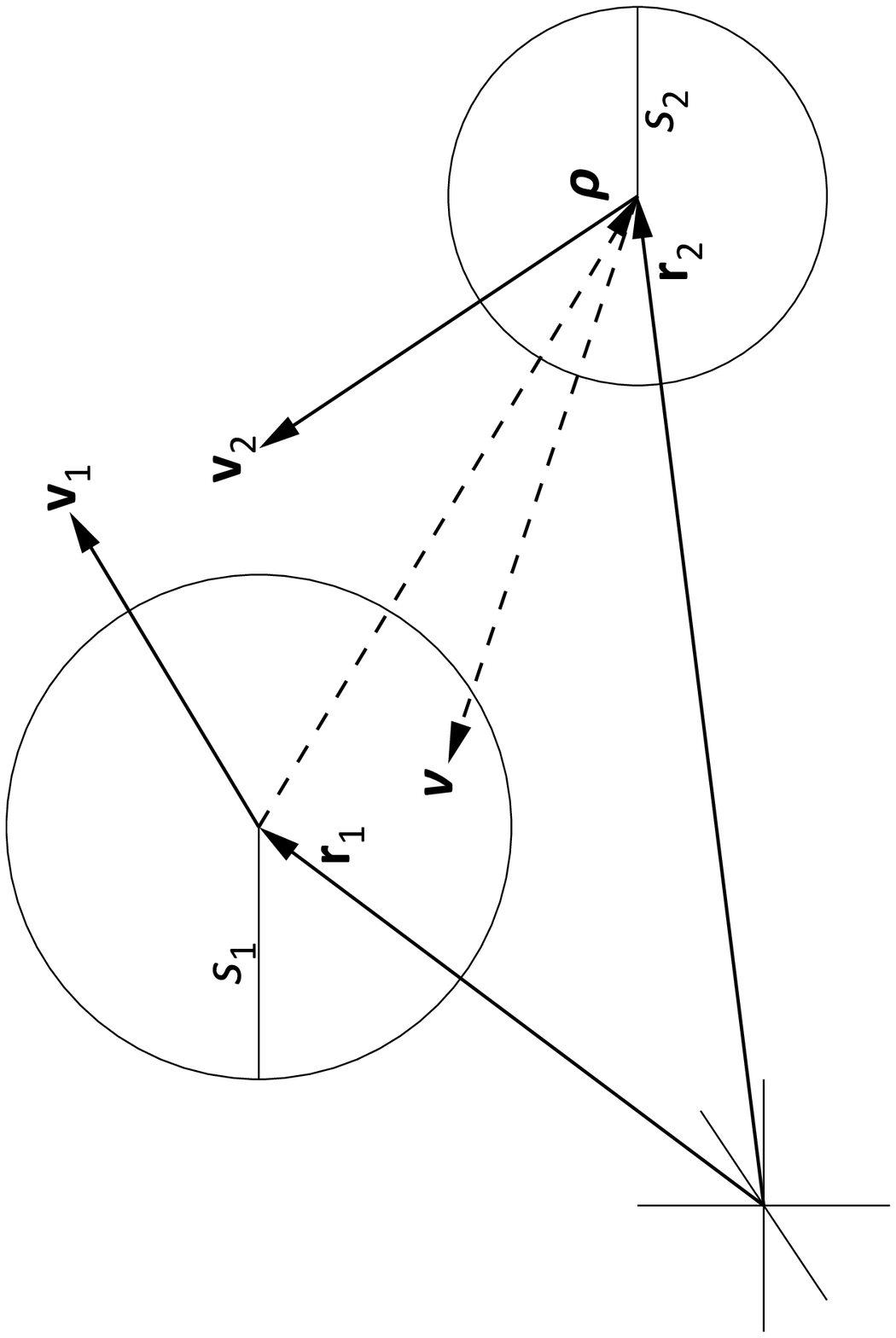}{\textwidth}
\putfig{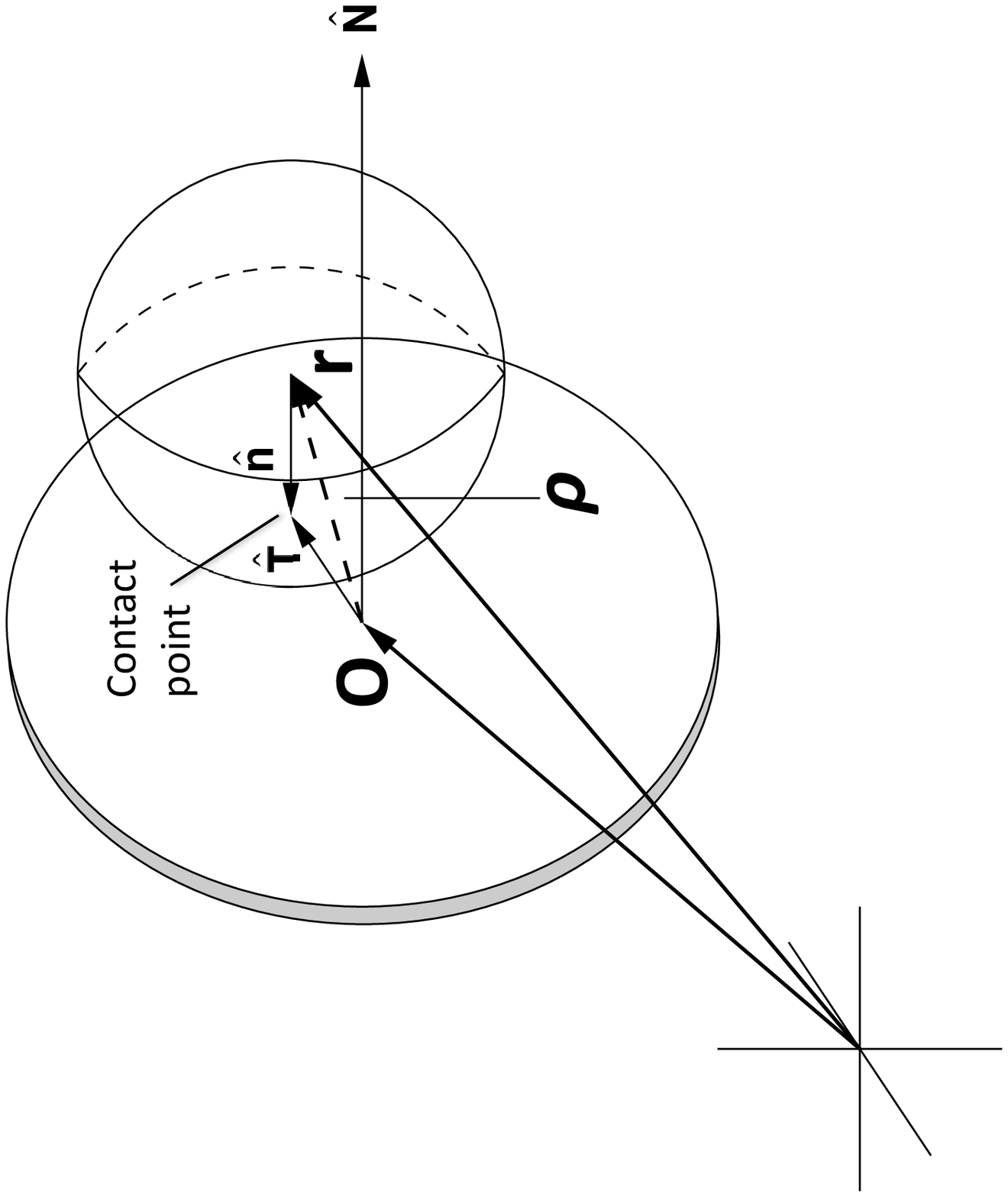}{\textwidth}
\putfig{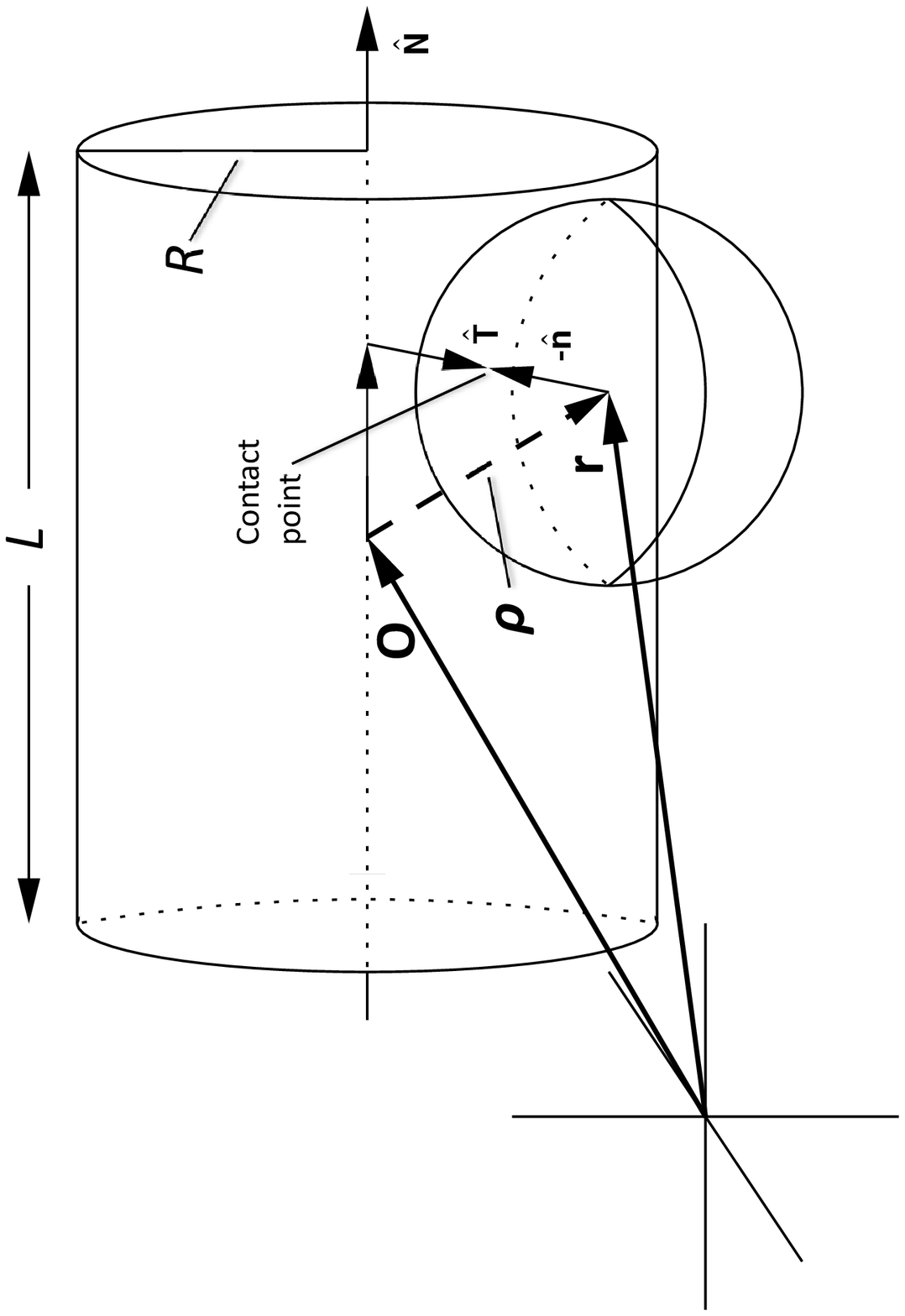}{\textwidth}
\putfig{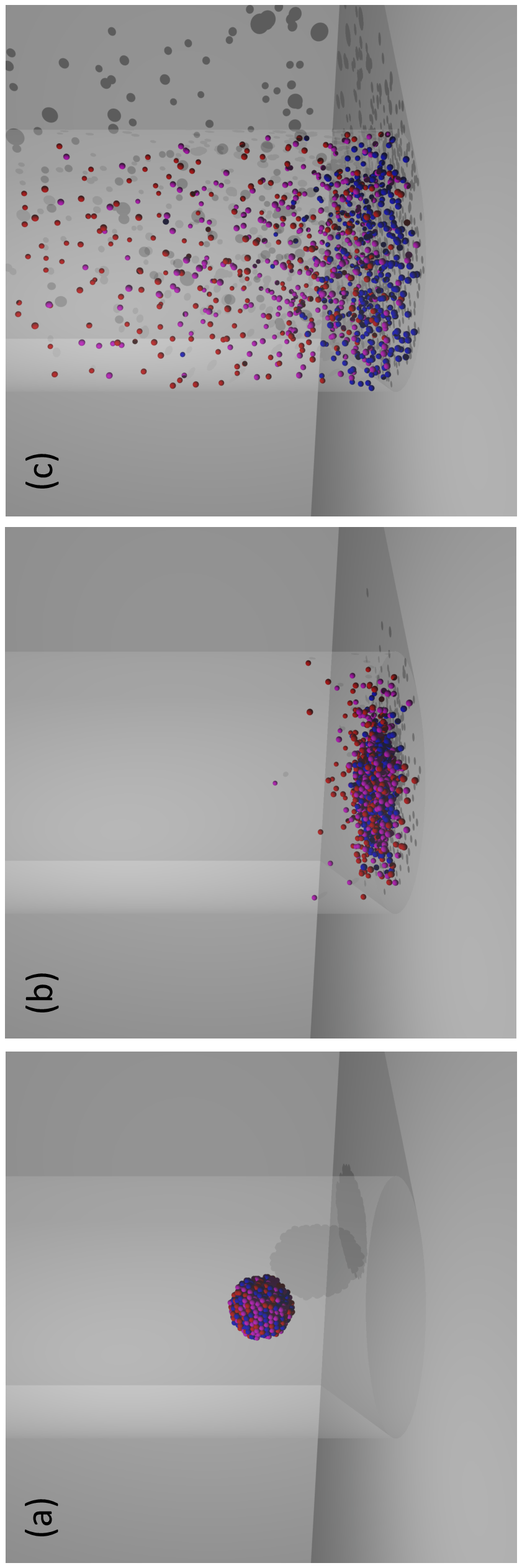}{\textwidth}
\putfig{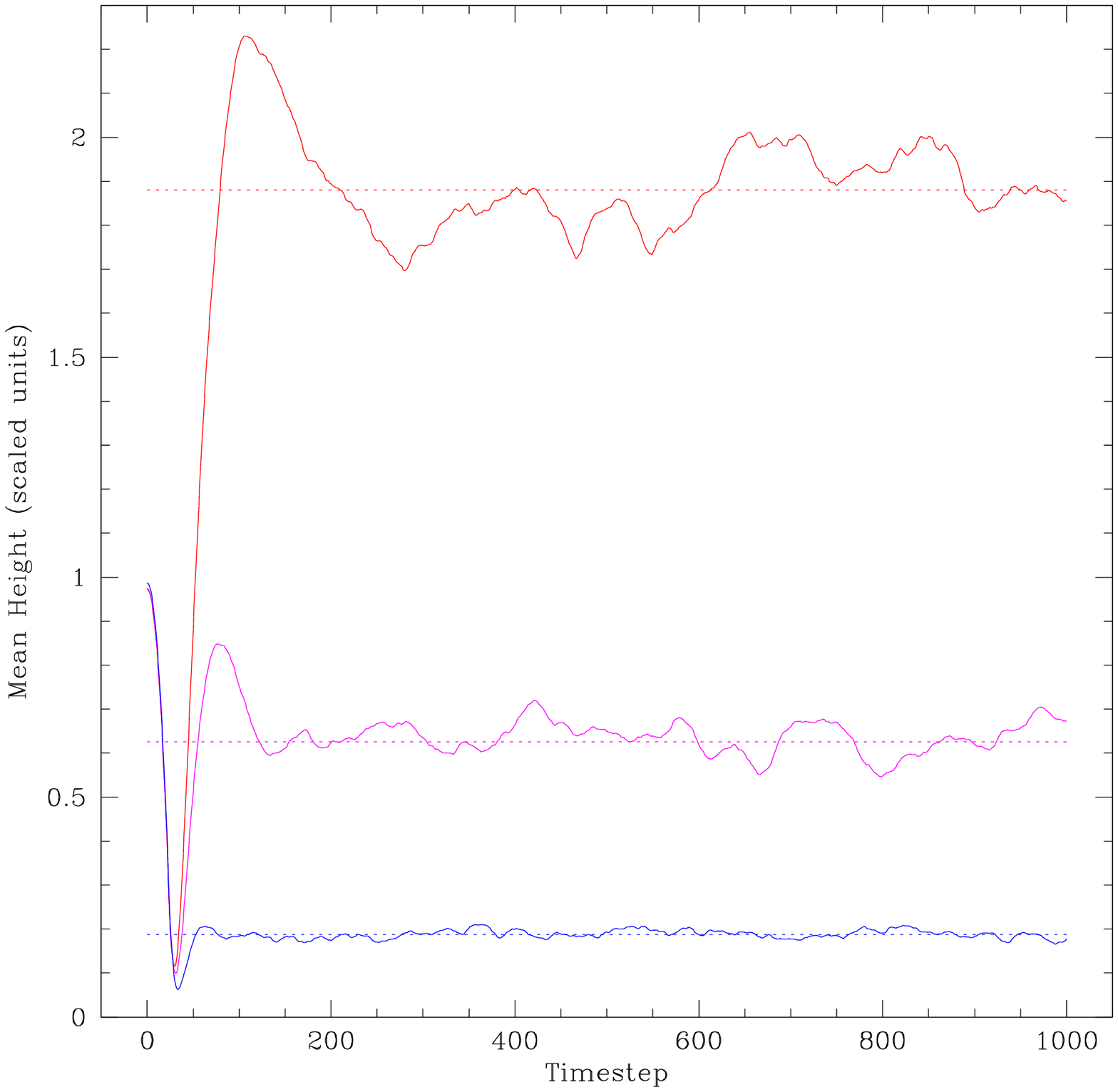}{\textwidth}
\putfig{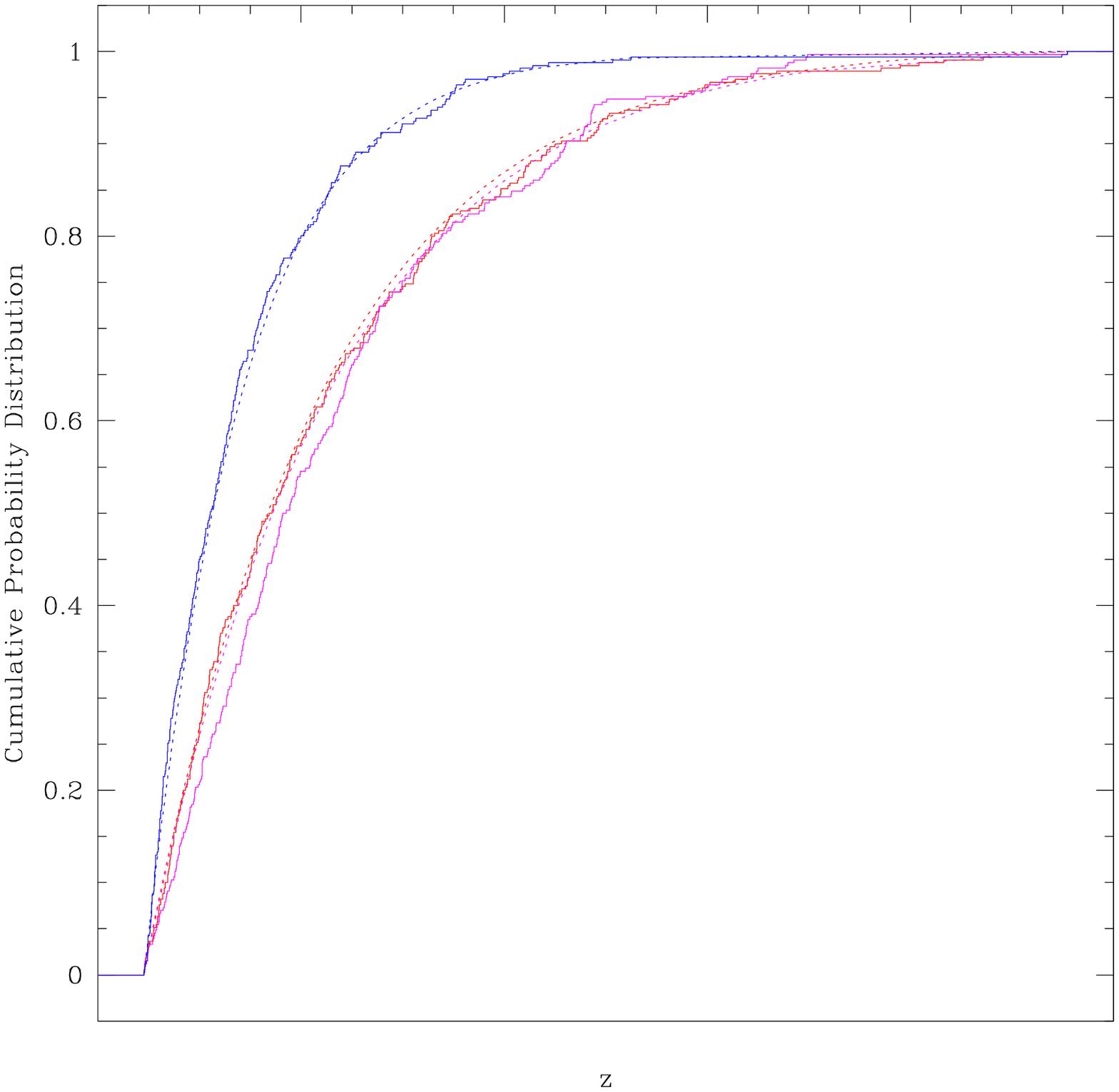}{\textwidth}
\putfig{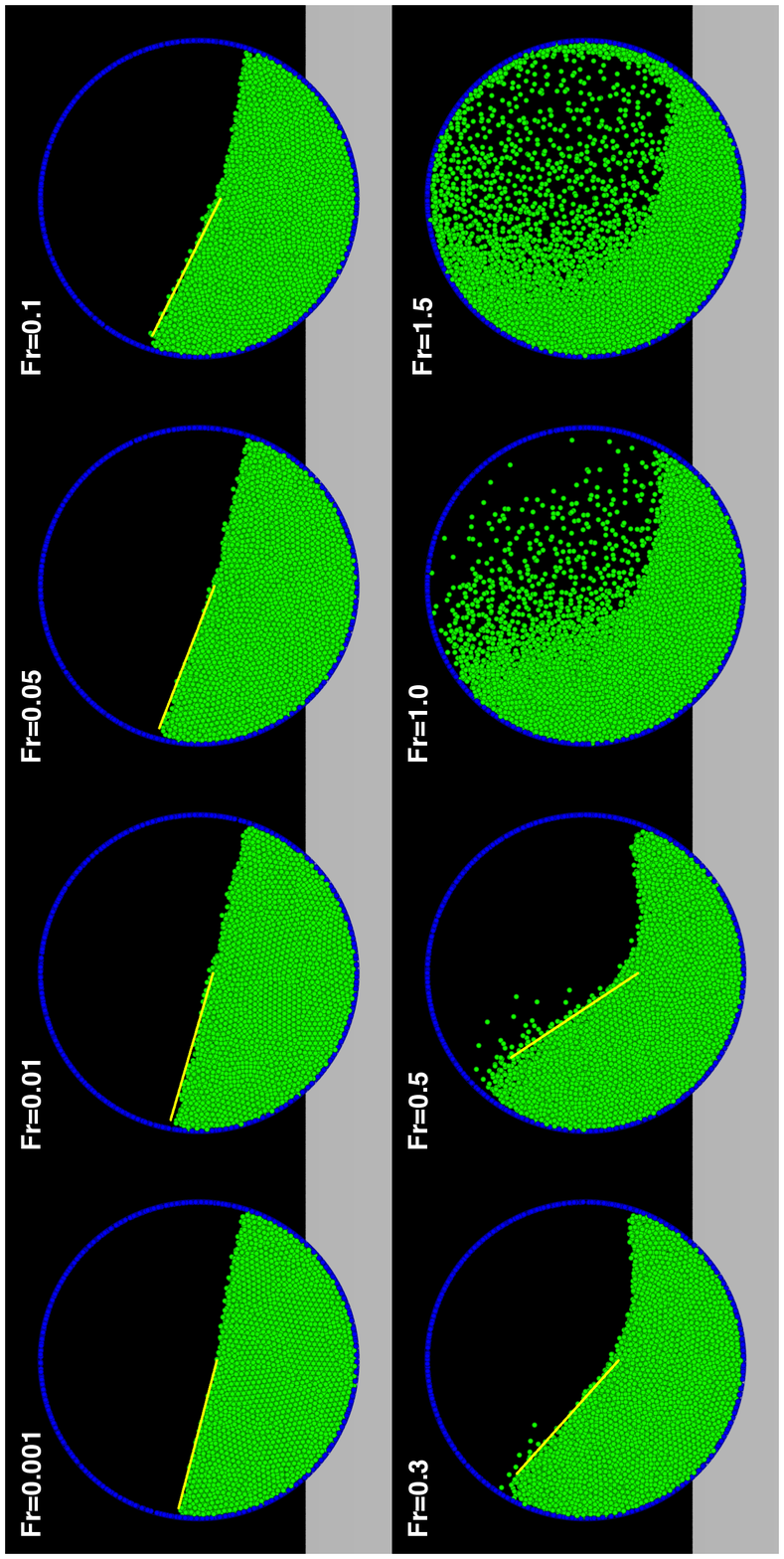}{\textwidth}
\putfig{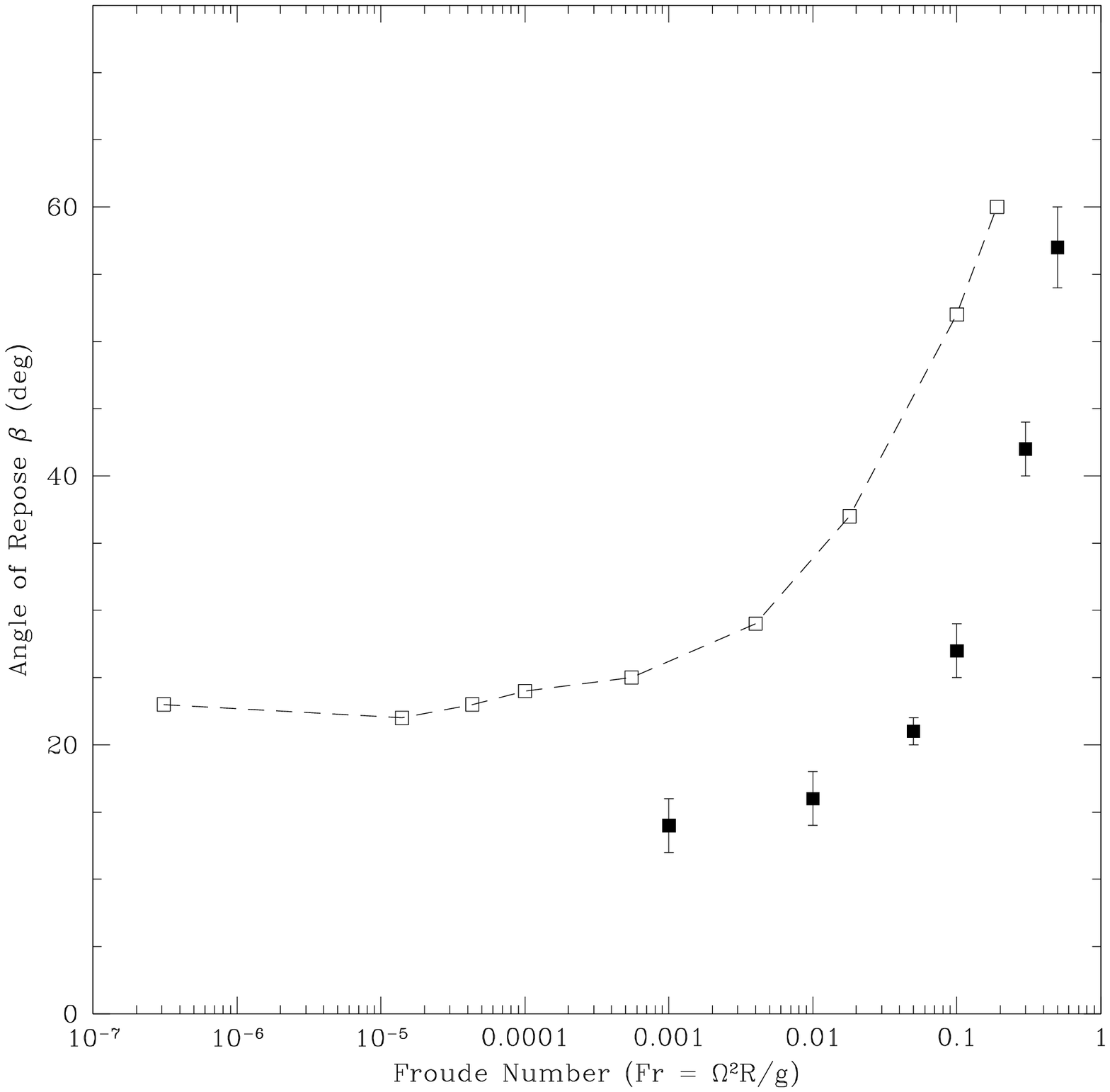}{\textwidth}

\end{document}